%% file: main.tex
\documentclass[a4paper,USenglish,cleveref,autoref,thm-restate]{lipics-v2021}

\usepackage{amsmath,amssymb}
\usepackage{booktabs}
\usepackage{graphicx}
\usepackage{microtype}
\usepackage{siunitx}
\usepackage{tikz}
\usetikzlibrary{arrows.meta, positioning}
\usepackage{makecell}

\graphicspath{{aft_paper/results/l1/}{aft_paper/results/l2_total/}{aft_paper/results/l2_multidim/}{aft_paper/results/sampling/}}

\title{Price Elasticity of Gas Demand on L1 and L2: Evidence from Ethereum and Arbitrum}

\titlerunning{Price Elasticity of Gas Demand on L1 and L2}

\author{Pranay Anchuri}{Offchain}{panchuri@offchain.io}{https://orcid.org/0000-0003-4377-5036}{}
\author{Akaki Mamageishvili}{Offchain}{amamageishvili@offchain.io}{https://orcid.org/0000-0003-2179-7867}{}

\authorrunning{Pranay Anchuri and Akaki Mamageishvili}



\category{}
\relatedversion{}
\supplement{}
\funding{}
\acknowledgements{}

\EventEditors{}
\EventNoEds{0}
\EventLongTitle{6th Conference on Advances in Financial Technologies (AFT 2026)}
\EventShortTitle{AFT 2026}
\EventAcronym{AFT}
\EventYear{2026}
\EventDate{}
\EventLocation{}
\EventLogo{}
\SeriesVolume{}
\ArticleNo{25}
\hideLIPIcs

\Copyright{Pranay Anchuri and Akaki Mamageishvili}

\usepackage{xcolor}
\newcommand{\rev}[1]{#1}

\ifdefined\finalversion
    \newcommand{\note}[3]{}
\else
    \renewcommand{\note}[3]{{\color{#2}\textbf{[#1: #3]}}}
\fi

\usepackage{xspace}
\usepackage[T1]{fontenc} 

\renewcommand{\eth}{Ethereum\xspace}
\newcommand{\arb}{Arbitrum\xspace}
\newcommand{\onefivefivenine}{EIP-1559\xspace}
\newcommand{\adv}{adversarial\xspace}
\newcommand{\Adv}{Adversarial\xspace}

\begin{document}
\nolinenumbers

\maketitle

\begin{abstract}
We estimate the causal price elasticity of gas demand on Ethereum mainnet (L1)
and Arbitrum One (L2), a quantity necessary for calibrating fee mechanism
simulations, evaluating resource pricing reforms, and explaining observed
usage patterns.
A two-way fixed effects panel regression instrumented by each wallet's own
lagged base fee removes the congestion-driven endogeneity that causes naive
regressions to substantially underestimate demand sensitivity.
On Ethereum mainnet (full year 2025),
the pooled IV elasticity is $-0.006^{***}$, near-inelastic: a $10\%$ fee increase
reduces total gas demand by approximately $0.06\%$.
On Arbitrum One (October 2025--April 2026),
the pooled IV elasticity is $-0.036^{**}$. Both chains are inelastic in the
aggregate, with L2 measurably more responsive than L1.
A per-resource decomposition of L2 demand reveals elasticities ranging from
modestly elastic computation ($-0.027^{*}$) to $-0.27^{***}$ for refunds,
with storage growth ($-0.15^{***}$) and calldata ($-0.06^{*}$) in between.
Behavioral clustering identifies always-on protocol wallets as near-inelastic
and high-volume operators as substantially more responsive, with cluster-level
elasticities up to roughly $6\times$ the pooled estimate.
These results establish an empirical foundation for downstream simulations
and for evaluating fee mechanism designs.
\end{abstract}

\section{Introduction}
\label{sec:intro}

On Ethereum, the base fee adjusts in each block in response to block utilization, targeting
50\% block capacity on average.
Despite the mechanism's theoretical properties being well-studied
\cite{roughgarden2021eip1559,roughgarden2021mechanism,chung2023,leonardos2021}, the empirical demand response
to fee changes remains poorly quantified.
Knowing how much users reduce their onchain resource consumption when fees rise is necessary
for multiple reasons. First, it allows calibrating simulations of the fee adjustment rule under different parameters. 
Second, evaluating potential throughput gains from multi-resource fee mechanisms requires demand elasticity estimates for each resource.
Third, it helps explain current usage patterns under different fee regimes. Fourth, it helps find the effects of repricing certain opcodes\footnote{The Ethereum ecosystem is considering increasing the gas target/limit while also increasing the gas cost of storage creation; Ethereum improvement proposal numbered 8037 (\url{https://eips.ethereum.org/EIPS/eip-8037}) is specifically proposed for this. Monad (also an EVM chain) increased the gas cost of storage creation at initialization time relative to Ethereum's default:~\url{https://category-labs.github.io/category-research/monad-initial-spec-proposal.pdf}}. 

\begin{figure}[t]
  \centering
  \includegraphics[width=0.72\linewidth]{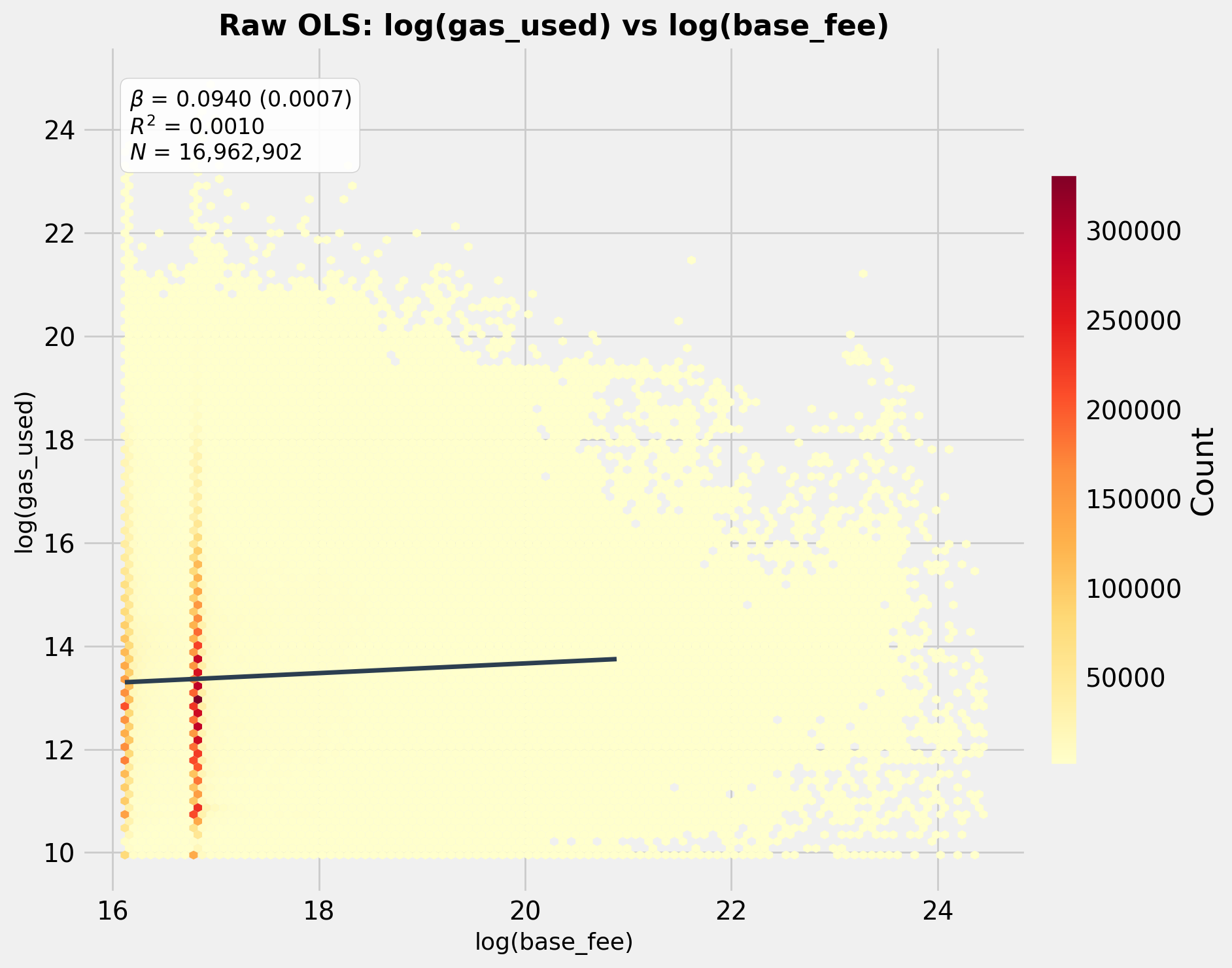}
  \caption{Raw pooled OLS of log gas usage on log base fee, \arb L2 panel
    ($N = \num{16939781}$ wallet-hour observations). The positive slope
    ($\hat{\beta} = +0.094$) contradicts the expected negative demand
    response, illustrating the upward bias induced by congestion-driven
    endogeneity.
    \rev{The two dense vertical bands at $\ln p \approx 16.1$ and $\approx 16.8$
    correspond to the \arb base fee floor levels of $0.01$~gwei and $0.02$~gwei
    respectively (the floor was raised from $0.01$ to $0.02$~gwei mid-sample),
    where the base fee is anchored when demand falls below the gas target.}}
  \label{fig:raw-scatter}
\end{figure}

Estimating demand elasticity naively is complicated by omitted variable bias.
Higher network demand drives the base fee up (through the EIP-1559) and gas usage up (through more users transacting), producing a spurious positive correlation between fees and gas even when isolated to wallet level.
Figure~\ref{fig:raw-scatter} illustrates this directly: a pooled OLS regression of log gas usage on log base fee across \num{16939781} \arb wallet-hour observations yields $\hat{\beta} = +0.094$, a positive coefficient that contradicts the expected negative demand response.
This is in contrast to a typical demand elasticity estimation scenario where the price change is mostly exogenous, e.g., due to a supply drop. In the context of Ethereum and its rollups, the target supply is fixed and the price (gas fees) follows the demand exactly, as described above.
Two-way fixed effects (FE: wallet and time) absorb level differences but still leave
within-cell congestion variation uncontrolled.
We address this residual endogeneity by using each wallet's own lagged base fee as an instrument: a fee level that is predetermined relative to current demand decisions and strongly correlated with the current fee through the autocorrelation built into the \onefivefivenine fee adjustment rule.

Gas is priced as a single unit on both chains, but the underlying operations vary substantially in their resource footprint: a storage-creating transaction consumes a different mix of compute, I/O, and state resources than a pure computation. Evaluating proposals for multi-dimensional resource pricing requires knowing which dimensions actually respond to fee signals. The same decomposition informs the demand impact of opcode repricing, where increasing the cost of one resource category may or may not shift usage depending on the price sensitivity of that dimension.

The pooled elasticity is an average across all wallet types on the chain: oracle updaters that submit at fixed intervals, DeFi operators that optimize gas costs, and MEV searchers that time activity to congestion. These populations differ fundamentally in their relationship to fee changes, and a single estimate characterizes none of them.

\paragraph{Contributions.}
\begin{enumerate}
  \item \textbf{Causal elasticity estimates for L1 and L2.}
        Using each wallet's own lagged base fee as an instrument, we obtain Instrumental Variables (IV) estimates of gas demand elasticity for Ethereum
        mainnet ($-0.006$) and Arbitrum One ($-0.036$).
  \item \textbf{Per-resource decomposition for L2.}
        We estimate separate elasticities for seven Arbitrum resource dimensions,
        revealing that state-creating and state-clearing operations are far more
        price-sensitive than computation or reads, which can be justified by their time-insensitivity.
  \item \textbf{Wallet-type heterogeneity.}
        K-means clustering on behavioral features identifies distinct user
        types with qualitatively different demand responses, from near-inelastic
        always-on protocol wallets to substantially elastic high-volume operators.
\end{enumerate}

The pooled elasticity estimates confirm near-inelastic demand on both chains: a $10\%$ fee increase reduces total gas demand by approximately $0.06\%$ on \eth and $0.36\%$ on \arb. However, per-resource decomposition reveals that this aggregate inelasticity masks substantial resource-level heterogeneity: refunds ($-0.27$) and storage growth ($-0.15$) are the most price-sensitive dimensions, while computation ($-0.027$) is near-inelastic. Wallet clustering further identifies congestion-timing wallets for which the lagged-fee instrument is invalid, and restricts causal interpretation to the remaining segments.

The remainder of the paper is organized as follows. Section~\ref{sec:background} reviews the EIP-1559 mechanism and L1/L2 fee regime differences. Section~\ref{sec:data} describes the Ethereum and Arbitrum datasets. Section~\ref{sec:econometrics} develops the econometric framework. Section~\ref{sec:clustering} presents the wallet clustering procedure. Sections~\ref{sec:l1} and~\ref{sec:l2-total} report the main elasticity results for L1 and L2 aggregate demand; Section~\ref{sec:l2-multidim} provides the per-resource decomposition. 
Section~\ref{sec:applications} discusses applications of elasticity analysis. Section~\ref{sec:discussion} discusses implications and limitations.

\section{Background}
\label{sec:background}

\subsection{The EIP-1559 Fee Mechanism}
Permissionless blockchains such as Ethereum need to control usage of their resources, to avoid long-lasting denial of service attacks and keep low hardware or bandwidth requirements for participating nodes in the network.
For this reason, the Ethereum chain introduced Ethereum improvement proposal (EIP), numbered 1559, which adjusts the fee for resource consumption.
There are different types of resources transactions consume, which are converted into gas units. Ethereum aims to operate on a target gas consumption. To achieve this goal, a simple rule is implemented: when the consumption of gas in the last block is larger than the target, the gas fee (in ETH) goes up; when the consumption of gas in the last block is lower than the target, the gas fee goes down; only when the gas usage is exactly equal to the target the gas fee stays the same. Gas fee changes should cause a short-term demand shift for the resources. Block sizes are usually upper bounded in terms of gas, commonly referred to as a block limit. The limit is set to two times the target. Recently, the target has moved up by a factor of 2 on Ethereum, from 15M to 30M, in an effort to scale the chain, with further increases expected.  

A key design goal of \onefivefivenine is fee predictability: the base fee is algorithmically calculated from the recent block utilization. This allows wallets to anticipate costs and plan resource consumption accordingly. This predictability both induces per-transaction optimization and deters participation during high-fee regimes. These are two distinct behavioral responses, and the former is the primary focus of this work.

\subsection{L1 vs.\ L2 Fee Regimes}
\rev{Major rollups adopt EIP-1559 or a variant of it to price their native resources
and manage congestion. L2 fees are typically lower than L1 fees for two reasons:
first, EIP-4844 expanded blob throughput so that L1 data availability costs for
rollups are now minimal; second, lower hardware and bandwidth requirements reduce
the marginal cost of an L2 block relative to an L1 block. Recent gas target
increases on L1 have compressed the gap further, making L1 fees relatively cheap
as well. The total fee incurred by an L2 end user consists of a native L2
congestion fee and an L1 data availability component; our analysis focuses on the
native L2 base fee.} 

Recently, Arbitrum updated its pricing regime, introducing a gas ladder\footnote{\href{https://forum.arbitrum.foundation/t/aip-raise-the-gas-target-min-l2-base-fee-implement-improvements-to-the-pricing-algorithm/30182}{Arbitrum gas-ladder AIP, Arbitrum Forum, 2025.}}. This pricing mechanism allows temporary outbursts of gas usage while keeping long-term targets fixed. 

An interesting difference between these two types of chains is that \eth does not have a lower bound on gas price; it can go as low as 1 wei ($10^{-18}$ ETH), while \arb does have it. Recently, with the pricing mechanism update, \arb increased it from 0.01 gwei to 0.02 gwei. A potential explanation for this is to deter spam transactions, the activity which is delegated to block builders on L1. Other major rollups enforce base fee floors that are substantially lower: OP Mainnet sets its minimum base fee to 0 wei (no effective floor), Base Mainnet to 0.005 gwei (\num{5000000} wei), and Scroll to approximately 0.00012 gwei (set via configuration rather than the protocol). Arbitrum's 0.02 gwei floor is the highest among these rollups, roughly $4\times$ Base's and orders of magnitude above the others.

The Arbitrum Nitro node is instrumented to track gas consumption across multiple resource dimensions (computation, storage reads and writes, storage growth, history growth, calldata, and refunds), attributing each opcode's cost to the corresponding category during execution. While the current Arbitrum fee mechanism charges a single gas price, the per-resource data provide a richer view of which operations are sensitive to fee changes, and form the basis for the per-resource demand decomposition in Section~\ref{sec:l2-multidim}.

Note that when the base fee is at the floor for extended periods, fee variation is compressed and the identifying variation exploited by the instrument becomes sparse. This constraint is specific to L2 estimation and does not arise on L1, where fee variation is continuous and unbounded from below.

\section{Data}
\label{sec:data}

\subsection{Ethereum L1}
\label{sec:data_l1}
The Ethereum dataset comprises all transactions executed on the \eth mainnet during 2025. Wallets are included in the study if they submitted at least 500 transactions during the year, ensuring sufficient within-wallet variation for the fixed effects estimation. \rev{After filtering on this threshold, the panel data contains \num{44457} wallets
observed over \num{8751} hourly periods, yielding a total of \num{35321371}
wallet-hour observations.}

\rev{For each wallet-hour, the dependent variable is the wallet's total gas used,
the independent variable is the wallet's own average base fee in that hour in gwei,
and the control is the number of transactions submitted.}

\rev{Both L1 and L2 panels are at hourly granularity. On \eth, we use daily time
fixed effects rather than hourly: the Ethereum base fee adjusts each block
(${\sim}12$ seconds), but the dominant fee variation is between days rather than
within a day. Daily fixed effects efficiently absorb aggregate day-level demand
shocks; the within-day, within-wallet hourly fee variation identifies $\beta$.}

\subsection{Arbitrum L2}
\label{sec:data_arb_total}

The \arb L2 dataset covers all \arb One transactions from October 2025 through
April 2026 (seven months). We use the same minimum threshold of \num{500} transactions to select the wallets. The resulting panel contains \num{23119} wallets observed over \num{5087} hourly periods, yielding a total of \num{16939781} wallet-hour observations.

In addition to aggregate gas usage, the \arb dataset records per-transaction
gas consumption across seven resource dimensions: \textit{computation},
\textit{historyGrowth}, \textit{storageAccessRead}, \textit{storageAccessWrite},
\textit{storageGrowth}, \textit{l2Calldata}, and \textit{refund}.
Table~\ref{tab:l2-dims} describes each dimension.
\rev{The dataset was constructed by instrumenting the Nitro execution engine to
attribute each opcode's gas cost to one of these resource categories during
transaction execution; the same opcode may map to different categories depending
on execution context (for example, an \texttt{SSTORE} counts toward
\textit{storageGrowth} on first write but \textit{storageAccessWrite} on updates).}
Computation is non-zero in every wallet-hour by construction; refund appears in
fewer than half, reflecting the discretionary nature of state-clearing operations.

\begin{table}[t]
  \centering
  \caption{L2 per-resource dimension coverage. Coverage is the fraction of
    total wallet-hours (\num{16939781}) where the resource consumption is non-zero.}
  \label{tab:l2-dims}
  \small
  \begin{tabular}{lp{0.30\linewidth}rrr}
    \toprule
    Dimension & Description & Wallets & Wallet-hour obs. & Coverage (\%) \\
    \midrule
    computation        & In-memory EVM execution                        & \num{23119} & \num{16939781} & 100.0 \\
    historyGrowth      & Logs and events (append-only)                  & \num{20052} & \num{13702494} &  80.9 \\
    storageAccessRead  & Reading existing state                         & \num{22652} & \num{16126735} &  95.2 \\
    storageAccessWrite & Updating existing state                        & \num{20883} & \num{13475351} &  79.6 \\
    storageGrowth      & Increasing persistent state size               & \num{18990} &  \num{8767695} &  51.8 \\
    l2Calldata         & Transaction calldata size on the child chain   & \num{20799} & \num{15782132} &  93.2 \\
    refund             & Gas refunded from storage-clearing operations  & \num{17770} &  \num{8103016} &  47.8 \\
    \bottomrule
  \end{tabular}
\end{table}

\subsection{Panel Data Construction}
\rev{Panels are constructed by aggregating raw transaction data to the
(wallet, hour) level for both chains. For each wallet-hour, all transactions
submitted by that wallet in the hour are aggregated to obtain total gas used and
transaction count, and the base fee is averaged across those transactions. Time
fixed effects are daily for \eth (365 day dummies over the hourly panel) and
hourly for \arb.} We exclude system and precompile addresses from both panels.

The instrument, the wallet's own lagged base fee, is computed by retrieving the most recent period during which the wallet was active and the average base fee during that period.

\section{Econometric Framework}
\label{sec:econometrics}
In this section, we develop the estimation strategy for the price elasticity of gas demand. We begin by showing why a naive regression of gas usage on the base fee produces a biased estimate, formalize the bias using the omitted-variable formula, and then build the solution in two steps.

\begin{itemize}
    \item Two-way fixed effects to absorb persistent wallet heterogeneity and time-dependent shocks.
    \item Instrumental variables to remove the residual endogeneity that fixed effects cannot eliminate.
\end{itemize}

\subsection{The Identification Problem}
\label{sec:bias}
The parameter of primary interest in this work is $\beta$, the price elasticity of gas demand: the percentage change in a wallet's gas consumption following a one-percent increase in the base fee. To isolate the source of bias, consider first a model with a single regressor,

\begin{equation}
  \label{eq:simple}
  \log g_i = \beta \log p_i + u_i,
\end{equation}
where $g_i$ is gas used, $p_i$ is the base fee, and $u_i \sim N(0,\sigma^2)$. We use a log-log specification throughout.

The ordinary least squares (OLS) estimator satisfies
\begin{equation}
  \hat{\beta}^{\mathrm{OLS}}
    = \frac{\widehat{\mathrm{Cov}}(\log p,\,\log g)}
           {\widehat{\mathrm{Var}}(\log p)},
\end{equation}
and converges in probability to
\begin{equation}
  \label{eq:plim}
  \operatorname{plim} \hat{\beta}^{\mathrm{OLS}}
    = \beta + \frac{\mathrm{Cov}(\log p,\, u)}{\mathrm{Var}(\log p)}.
\end{equation}
OLS is consistent if and only if $\mathrm{Cov}(\log p,\, u) = 0$.

In our setting, this condition fails because of (latent) aggregate demand pressure, or congestion. Denote by $D_t$ the aggregate level of demand in the time period $t$. Under the \onefivefivenine adjustment rule, the base fee for each block is adjusted by a factor that is proportional to the excess utilization of the preceding block. Therefore, congestion raises the fee: $\mathrm{Cov}(\log p,\, D_t) > 0$. At the same time, a high-congestion period reflects users submitting more transactions, which raises the usage: $\mathrm{Cov}(\log g,\, D_t) > 0$. The term $D_t$ is part of the error $u$ in Equation~\eqref{eq:simple} and these two conditions combine to give $\mathrm{Cov}(\log p,\, u) > 0$. Therefore, the bias term in Equation~\eqref{eq:plim} is positive. OLS overestimates the true elasticity $\beta$, biasing it toward zero or even reversing its sign. This is a structural bias and cannot be eliminated by increasing the sample size.

\subsection{Two-Way Fixed Effects Model}
\label{sec:fe}
The congestion described in Section~\ref{sec:bias} operates on two easily separable levels. First, wallets differ persistently: a high-frequency trading bot and a casual user differ in their gas consumption across all time periods for reasons unrelated to the base fee. Second, the network-level congestion shocks are common across all wallets within a given period. For example, a fee spike during an NFT drop affects all wallets simultaneously. Wallet fixed effects absorb the first source and the time fixed effects absorb the second.

\paragraph{Demand equation}
The demand equation for wallet $i$ in time period $t$ is
\begin{equation}
  \label{eq:structural}
  \log g_{it} = \alpha_i + \gamma_t + \beta \log p_{it} + \phi \log n_{it} + u_{it},
\end{equation}
where $p_{it}$ is wallet $i$'s average base fee in period $t$, $n_{it}$ is
the number of transactions submitted (a control variable), and
$u_{it}$ is the residual. The term $\alpha_i$ absorbs all the time-invariant heterogeneity between wallets. These include bot versus human, typical gas usage per transaction, etc. The term $\gamma_t$ absorbs all shocks at the period-level experienced by all wallets. These include network congestion level, time of day, day of week, and macro demand events. We also include the transaction count $n_{it}$ because a wallet with more transactions mechanically consumes more gas. Controlling for $n_{it}$ isolates the response in usage to the fee. 

The control for $n_{it}$ reflects a deliberate choice about the margin being estimated. The total gas $g_{it}$ combines two distinct decisions made by the wallets: how many transactions to submit in a period $t$ and how much gas to spend in each transaction. The extensive margin, transaction count, captures the participation decisions, whereas the intensive margin, gas used in a transaction, captures things such as reducing gas usage or substituting toward cheaper resources. By controlling for $n_{it}$, we partial out the extensive margin and identify $\beta$ as the intensive margin elasticity. The latter is directly relevant in evaluating the efficiency of different fee mechanisms, since fee mechanisms should not merely deter activity but enable wallets to optimize their resource usage. Because $n_{it}$ is fee-responsive, conditioning on it isolates the per-transaction (intensive-margin) response rather than the total effect, which is the margin most relevant to mechanism efficiency.%
\footnote{Omitting $\log n_{it}$ from both stages reverses the sign of the estimate
($+0.052$ on \eth, $+1.52$ on \arb): high-fee periods attract wallets that submit
more and heavier transactions, so total gas co-moves positively with fees when
transaction count is uncontrolled. Conditioning on $n_{it}$ isolates the
per-transaction response.}
\rev{We condition on $n_{it}$ rather than instrument for it because our estimand
is the intensive-margin elasticity: instrumenting for $n_{it}$ would recover the
total elasticity, a different quantity that conflates fee-driven participation
changes with per-transaction resource adjustments.}

After absorbing the terms $\alpha_i$ and $\gamma_t$, the elasticity parameter $\beta$ is identified from within-wallet and within-period variation. More specifically, $\beta$ measures how a wallet $i$'s
gas usage changes when the fee deviates from both the wallet's own time-average
and the period's cross-wallet average.

\paragraph{Two-way demeaning}
Estimating $\beta$ from Equation~\eqref{eq:structural} requires either one-hot encoding the fixed effects or demeaning the outcome and the regressors.
Let $\bar{y}_{i\cdot}$ denote the wallet time-mean, $\bar{y}_{\cdot t}$ the
period cross-wallet mean, and $\bar{y}_{\cdot\cdot}$ the overall mean.
The demeaned outcome is
\begin{equation}
  \label{eq:demean}
  \tilde{y}_{it}
    = y_{it} - \bar{y}_{i\cdot} - \bar{y}_{\cdot t} + \bar{y}_{\cdot\cdot},
\end{equation}
and the same transformation is applied to the regressors $\log p_{it}$ and $\log n_{it}$.

By the Frisch--Waugh--Lovell theorem~\cite{frisch1933partial,lovell1963seasonal}, OLS on the demeaned model is identical to including dummy variables for $\alpha_i$ and $\gamma_t$. We demean all variables by subtracting wallet and period means in alternation until convergence. This process terminates in fewer than 30 iterations in all panel datasets used in this study; the post-estimation fixed-effects recovery described in the full version of the paper
applies the same algorithm to the full residuals and converges within 500 iterations.

\paragraph{Time granularity} \rev{Both analyses use hourly panels; the granularity
of $\gamma_t$ differs. On Ethereum mainnet, the base fee adjusts by at most
$\pm 12.5\%$ per block (approximately 12~seconds), but the dominant variation is
between days rather than within a day. We therefore use daily time fixed effects:
$\gamma_t$ absorbs day-level aggregate demand shocks, and $\beta$ is identified
from within-day, within-wallet hourly fee deviations. On \arb, hourly fixed
effects are used because the L2 base fee adjusts on an hourly schedule.}

\paragraph{Residual endogeneity} The above two-way demeaning substantially attenuates the bias. The FE-OLS estimate of $\beta$ is $-0.007$ on \eth and $-0.016$ on \arb, both negative and consistent with downward-sloping demand. Within-period demand shocks remain in $\tilde{u}_{it}$ even after demeaning. For example, a wallet that transacts more in period $t$ because of within-period congestion
that also raised the fee in that period, contributes a positive correlation
between $\tilde{p}_{it}$ and $\tilde{u}_{it}$. The two-way FE removes between-period congestion but not within-period endogeneity. This motivates the instrumental variables approach below.

\subsection{Instrumental Variables}
\label{sec:iv}

Figure~\ref{fig:dag} illustrates the causal structure of the model.
Congestion $u_{it}$ drives both the base fee $p_{it}$ and gas usage $g_{it}$
through separate channels, creating the endogeneity that results in an
inconsistent estimate of the elasticity $\beta$.
An instrumental variable $z_{it}$ that shifts the fee but has no direct
effect on gas demand breaks this linkage.

\input{aft_paper/paper/causal_fig.tex}

For each wallet $i$ in period $t$, the instrument is the wallet's own average base fee in its immediately preceding active period:
\begin{equation}
  \label{eq:instrument}
  z_{it} = \log p_{i,t-1},
\end{equation}
where $t-1$ denotes the most recent period before $t$ in which wallet $i$
submitted at least one transaction.

For $z_{it}$ to be a valid instrument, the properties of \textit{Relevance} and \textit{Exclusion} must be established.

\paragraph{Relevance} The instrument must be correlated with the current fee:
$\mathrm{Cov}(z_{it},\, p_{it}) \neq 0$.
The base fee exhibits strong autocorrelation because \onefivefivenine adjusts it
continuously in response to prior block utilization, so the fee in one period is a strong predictor of the fee in the next.
The first-stage partial $F$-statistic confirms this correlation empirically. Results reported in Sections~\ref{sec:l1} and~\ref{sec:l2-total} far exceed the conventional threshold of~10 for reliable inference.

\paragraph{Exclusion Restriction} The validity of the instrument also relies on the assumption that it is uncorrelated with the error $u_{it}$. The lagged base fee is \textit{predetermined}, as it is a function of gas utilization in periods preceding $t$. Therefore, it is established before a wallet makes a marginal transaction decision at time $t$. Moreover, the two-way FE controls for both the time-invariant wallet characteristics and time-specific demand shocks. Under these conditions, the lagged fee affects the current usage only through its influence on the current base fee.

However, the exclusion restriction property is violated by wallets that strategically time their activity with congestion cycles. If a wallet's activity is sparse then the lagged fee $z_{it}$ represents the peak of a prior congestion. Now, if the congestion is serially correlated then the instrument $z_{it}$ becomes an informative signal of the current error term $u_{it}$. This leads to a correlation between the instrument and the latent demand shocks.

To mitigate the bias introduced by this channel, we identify such congestion sensitive wallets based on the proportion of wallet's activity during periods where the base fee exceeds the median hourly base fee. Wallets with an anomalously high fraction are the ones that are disproportionately active during spikes. For these wallets, the lagged fee is invalid as an instrument. \rev{We identify this as cluster C5 (congestion-active, Section~\ref{sec:clustering}): the median C5 wallet conducts 75.4\% of its active hours in above-median-fee periods (IQR: 40.0--87.9\%), compared with 27.0--48.1\% for the five retained clusters (Table~\ref{tab:exclusion-timing}). We therefore restrict causal interpretation to the five clusters whose timing patterns are consistent with the exclusion restriction.}

\begin{table}[t]
  \caption{Fraction of active hours in above-median-fee periods, by cluster (L2).}
  \label{tab:exclusion-timing}
  \centering\small
  \begin{tabular}{llrrrr}
    \toprule
    Cluster & Label & $n$ wallets & Median & P25 & P75 \\
    \midrule
    C0 & Hi-vol burst              & 2{,}608 & 0.367 & 0.268 & 0.474 \\
    C1 & Long-running              & 3{,}279 & 0.481 & 0.328 & 0.596 \\
    C2 & Low-intensity             & 5{,}818 & 0.270 & 0.239 & 0.323 \\
    C3 & Fee-avoiding              & 4{,}253 & 0.270 & 0.224 & 0.329 \\
    C4 & Variable gas              &   697   & 0.302 & 0.248 & 0.363 \\
    C5 & Cong-active$^{\dagger}$   & 6{,}466 & 0.754 & 0.400 & 0.879 \\
    \bottomrule
  \end{tabular}
  \par\smallskip
  \raggedright\footnotesize $^{\dagger}$Excluded from IV estimation (Section~\ref{sec:iv}).
\end{table}

\subsection{Calculating $\beta$ using Two-Stage Least Squares (2SLS)}
\label{sec:2sls}
Building on the discussion in this section, we now describe how we estimate the elasticity $\beta$. Prior to estimation, all variables are two-way demeaned as described in Section~\ref{sec:fe}.

In the first stage, the demeaned fee is regressed on the demeaned instrument and the demeaned transaction count (control):

\begin{equation}
  \label{eq:first-stage}
  \tilde{p}_{it} = \pi\,\tilde{p}_{i,t-1} + \delta\,\tilde{n}_{it} + v_{it}.
\end{equation}
Here $\tilde{p}_{i,t-1}$ is shorthand for the two-way demeaned instrument
$\tilde{z}_{it}$, where $z_{it} = \log p_{i,t-1}$.
The fitted values $\hat{p}_{it}$ capture only the variation in the fee that
is driven by the instrument.

In the second stage, the fee is replaced with the fitted values from the first stage.

\begin{equation}
  \label{eq:second-stage}
  \tilde{g}_{it} = \beta\,\hat{p}_{it} + \phi\,\tilde{n}_{it} + \varepsilon_{it}.
\end{equation}

With a single endogenous regressor (base fee) and a single instrument, the estimate from 2SLS reduces to 
\begin{equation}
  \label{eq:wald}
  \hat{\beta}^{\mathrm{IV}}
    = \frac{\widehat{\mathrm{Cov}}(z,\,\tilde{g})}
           {\widehat{\mathrm{Cov}}(z,\,\tilde{p})}.
\end{equation}

\paragraph{Interpretation}
The 2SLS estimator in Equation~\eqref{eq:wald} identifies the local average treatment effect (LATE). It is the average demand response for wallets whose fee in period $t$ is affected by the lagged fee. This is the subpopulation of wallets that respond to prior price levels when deciding the current gas usage. Wallets that execute on a fixed schedule regardless of the base fee are not shifted by the instrument and consequently have negligible contribution to the estimate.

\paragraph{Preview of main results}
\rev{Applying this framework to the full panels, the pooled IV elasticity is
$-0.006^{***}$ (SE~$= 0.0005$) on \eth and $-0.036^{**}$ (SE~$= 0.014$) on \arb.
Both estimates are near-inelastic and are detailed in Sections~\ref{sec:l1} and
\ref{sec:l2-total}. Section~\ref{sec:clustering} and Section~\ref{sec:l2-multidim}
decompose these pooled estimates by behavioral cluster and resource dimension,
respectively.}

\section{Wallet Clustering}
\label{sec:clustering}

The IV $\beta$ estimate from Section~\ref{sec:econometrics} is a precisely identified quantity, but it is a weighted average of demand responses across all wallet types in the data. In general-purpose blockchain networks like \eth and \arb, wallets span a wide range of behavioral regimes: oracle updaters, arbitrageurs, DeFi operators, and retail users. These wallets differ not only in their average gas consumption but also in how they respond to fee changes. An oracle contract that must submit a transaction at regular intervals to maintain the on-chain state faces no meaningful price sensitivity. At the same time, a high-volume arbitrageur may substitute toward lower fee periods. Pooling all wallet types into a single regression Equation~\eqref{eq:structural} produces an average elasticity estimate that does not characterize any particular wallet class.

Clustering the wallets into behaviorally homogeneous groups prior to IV estimation recovers within-cluster elasticities that are interpretable as the demand response of a coherent user type. It also serves a diagnostic role in our analysis: wallets that activate disproportionately during congestion spikes are localized in a single cluster identifiable by their high-fee fraction, as discussed in Section~\ref{sec:iv}.

\subsection{Clustering Procedure}
Wallets are partitioned using $k$-means clustering based on a set of behavioral features computed from each wallet's transaction history over the entire sample period.  Before clustering, we also standardize the features to zero mean and unit variance. We select the number of clusters $k$ with the elbow method: the within-cluster sum of squared distances is computed for $k = 2, \ldots, 12$, and $k$ is chosen at the point of diminishing returns.

Elbow plots for all three analyses are shown in the full version of the paper.
all three exhibit a clear kink at $k = 6$.

\subsection{Features}
\label{sec:clustering-features}

Table~\ref{tab:features} lists all features used for clustering. The five behavioral features are used in both the total-gas and per-resource analyses. For the resource level analysis, we add four resource composition features. Each of these features measures the fraction of a wallet's total gas attributable to a given resource dimension averaged over the entire time period.

The remaining dimensions (storageAccessWrite, historyGrowth, refund) are excluded from the clustering features as their variation is already captured by the included features.

\input{aft_paper/paper/feature_tab.tex}

\subsection{Cluster Profiles}
\label{sec:clustering-profiles}

\subsubsection{L1 clusters}
\label{sec:l1_cluster_profiles}
\begin{figure}[t]
  \centering
  \includegraphics[width=0.80\linewidth]{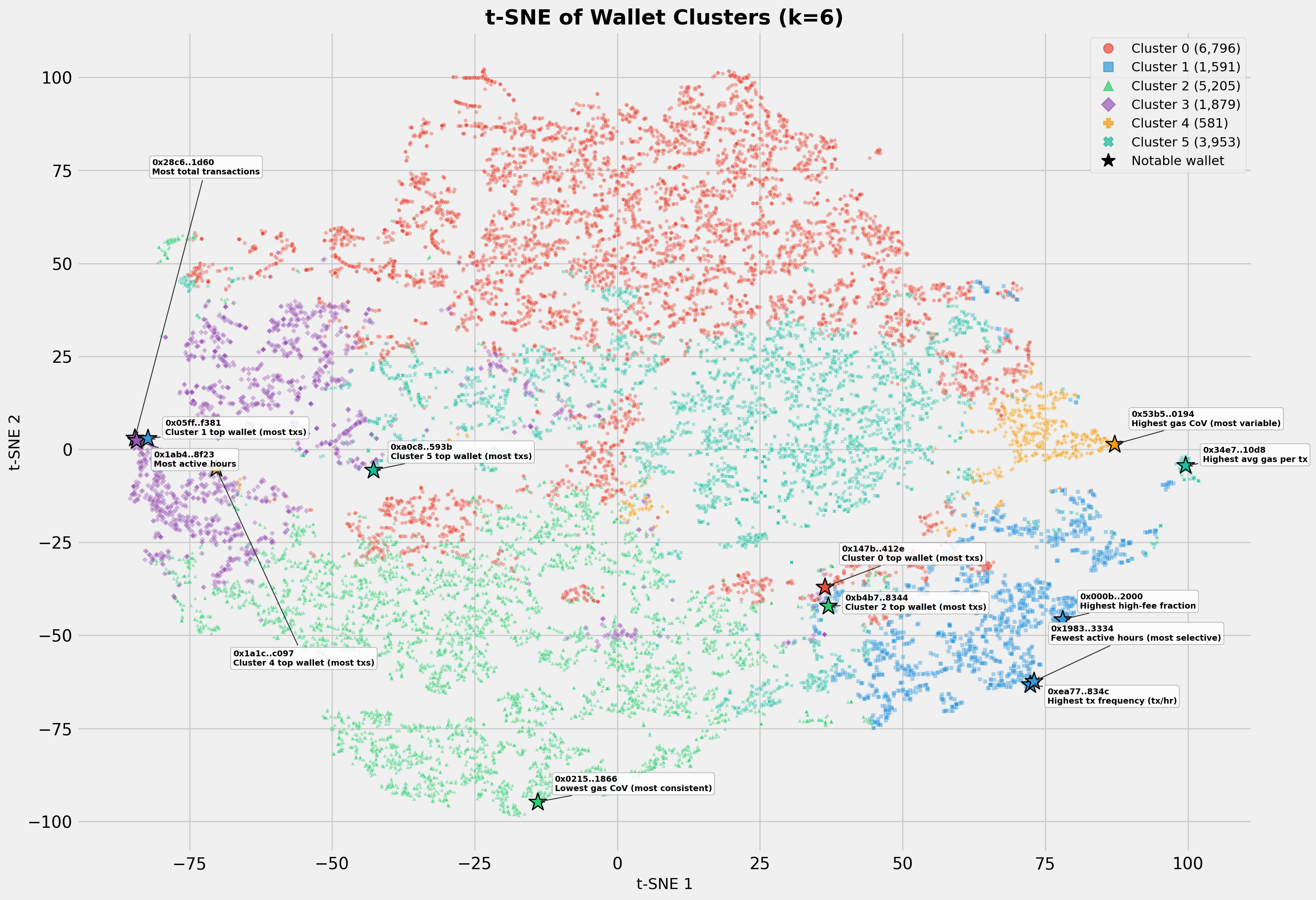}
  \caption{t-SNE projection of L1 wallet feature vectors ($k = 6$,
    five behavioral features). Each color represents one cluster.
    Clusters occupy well-separated regions, confirming meaningful
    behavioral separation.}
  \label{fig:tsne-l1}
\end{figure}

Table~\ref{tab:clusters-l1} reports the six L1 clusters, ordered by size.
Figure~\ref{fig:tsne-l1} shows a t-SNE projection of the five-dimensional feature space, confirming that the resulting clusters occupy geometrically distinct regions.

The clusters reflect qualitatively different wallet types. C3 is the most persistently active, with a median of 3309 active hours (38\% of all hours in the year). C1 has the highest frequency (25 transactions per active hour) but operates in short bursts (53 median active hours) and generally avoids high-fee periods (hff = 0.34). These wallets are likely to be opportunistic and sensitive to the base fee. C0 is highly active during high-fee periods suggesting price insensitivity or time-sensitive applications. C4 is the smallest cluster but exhibits the highest gas coefficient of variation, indicating wallets that alternate between simple and complex transactions. C5 is similar to C1 with respect to high-fee fraction but uses $3.5\times$ more gas per transaction.

\begin{table}[t]
  \centering
  \caption{L1 (Ethereum) wallet clusters ($k = 6$, full year 2025).
    Medians reported for each feature.
    tx\_freq: transactions per active hour;
    gas/tx: average gas per transaction (thousands);
    act.~hrs: active hours in sample;
    gas CoV: coefficient of variation of gas per transaction;
    hff: high-fee fraction.}
  \label{tab:clusters-l1}
  \small
  \begin{tabular}{clrrrrrr}
    \toprule
    Cl. & Wallets (\%) &
    tx\_freq & gas/tx (K) & act.~hrs & gas CoV & hff \\
    \midrule
    C0 & 15,135 (34.0\%) & 2.52 &  130 &   330 & 0.92 & 0.69 \\
    C1 &  3,609 ( 8.1\%) & 25.1 &   50 &    53 & 0.96 & 0.34 \\
    C2 & 11,625 (26.1\%) & 1.54 &   35 &   543 & 0.68 & 0.50 \\
    C3 &  4,123 ( 9.3\%) & 2.93 &  114 & 3,309 & 0.82 & 0.50 \\
    C4 &  1,280 ( 2.9\%) & 3.04 &  106 &   368 & 3.33 & 0.53 \\
    C5 &  8,685 (19.5\%) & 2.86 &  171 &   303 & 1.01 & 0.34 \\
    \bottomrule
  \end{tabular}
\end{table}

\subsubsection{L2 clusters (total gas).}
\label{sec:l2_cluster_profile}
\begin{figure}[t]
  \centering
  \includegraphics[width=0.80\linewidth]{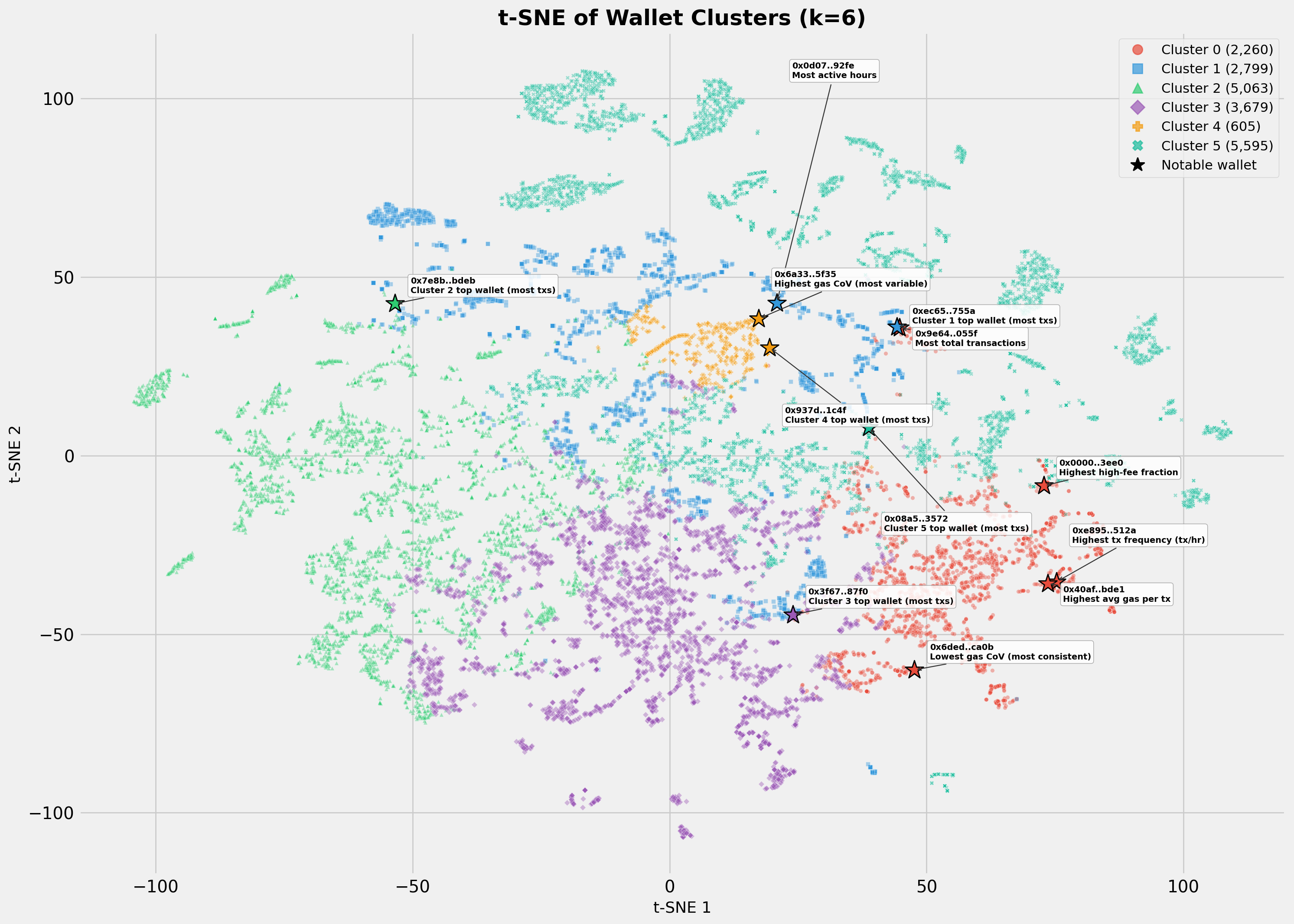}
  \caption{t-SNE projection of L2 total-gas wallet feature vectors
    ($k = 6$, five behavioral features). Clusters are well-separated.
    C5 (high-fee fraction 0.90, upper right) forms a tight distinct
    region.}
  \label{fig:tsne-l2}
\end{figure}

Table~\ref{tab:clusters-l2} reports the six L2 clusters.
Figure~\ref{fig:tsne-l2} shows the corresponding t-SNE projection.

\begin{table}[t]
  \centering
  \caption{L2 (Arbitrum) total-gas wallet clusters ($k = 6$,
    October 2025--April 2026).
    Medians reported. gas/hr: mean gas used per active hour (millions).
    Other columns as in Table~\ref{tab:clusters-l1}.}
  \label{tab:clusters-l2}
  \small
  \begin{tabular}{clrrrrrr}
    \toprule
    Cl. & Wallets (\%) &
    tx\_freq & gas/hr (M) & act.~hrs & gas CoV & hff \\
    \midrule
    C0 & 2,608 (11.3\%) & 58.74 & 16.87 &    24 & 1.02 & 0.53 \\
    C1 & 3,279 (14.2\%) &  8.43 &  2.14 & 2,858 & 1.25 & 0.51 \\
    C2 & 5,818 (25.2\%) &  2.31 &  0.17 &   532 & 0.97 & 0.54 \\
    C3 & 4,253 (18.4\%) &  5.24 &  1.11 &   166 & 0.90 & 0.34 \\
    C4 &   697 ( 3.0\%) &  3.42 &  0.51 &   898 & 4.42 & 0.48 \\
    C5 & 6,466 (28.0\%) & 16.81 &  0.90 &   117 & 1.48 & 0.90 \\
    \bottomrule
  \end{tabular}
\end{table}

C0 is the most active cluster by transaction rate (59 transactions per active hour) and with the highest mean hourly gas consumption (16.9M gas per hour) but with a very short median activity. C1 combines modest transaction activity with a very high number of active hours. C2 is the biggest cluster and has both low transaction frequency and gas usage. C3 has the lowest high-fee fraction, indicating wallets that are sensitive to the base fee. C4 has very high gas coefficient of variation. C5 is the cluster with highest high-fee fraction. The wallets transact predominantly in the most expensive gas fee regimes. As discussed in Section~\ref{sec:iv}, the exclusion property is implausible for this cluster.

\begin{figure}[t]
  \centering
  \includegraphics[width=0.80\linewidth]{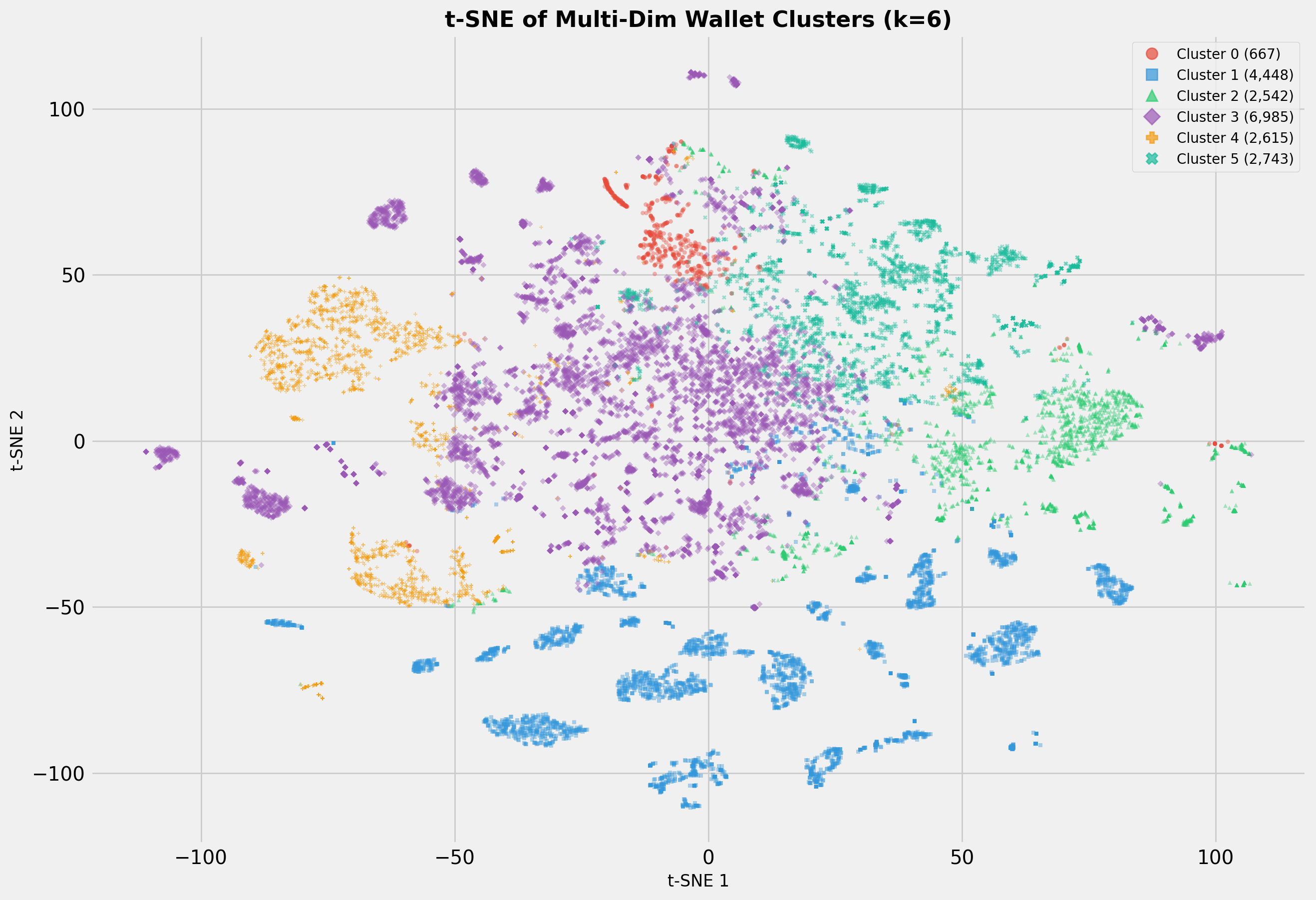}
  \caption{t-SNE projection of L2 per-resource wallet feature vectors
    ($k = 6$, nine features). Resource composition features produce
    additional separation not visible in the behavioral-only projection.}
  \label{fig:tsne-multidim}
\end{figure}

\paragraph{L2 clusters (per-resource).}
Table~\ref{tab:clusters-l2-multidim} reports the six clusters from the per-resource analysis, which uses nine features. Figure~\ref{fig:tsne-multidim} shows the t-SNE projection. The resource composition features listed in the final column are useful in interpreting the clusters.

\begin{table}[t]
  \centering
  \caption{L2 per-resource wallet clusters ($k = 6$,
    October 2025--April 2026, nine features).
    Medians reported. gas/hr: mean gas per active hour (thousands).
    Dominant dim: resource dimension with the highest median fraction.}
  \label{tab:clusters-l2-multidim}
  \small
  \begin{tabular}{clrrrrrl}
    \toprule
    Cl. & Wallets (\%) &
    tx\_freq & gas/hr (K) & gas CoV & hff & Dominant dim. \\
    \midrule
    C0 &   774 ( 3.3\%) &  3.76 &    608 & 4.21 & 0.49 & balanced (comp 42\%, read 26\%) \\
    C1 & 5,132 (22.2\%) & 17.47 &    845 & 1.53 & 0.91 & comp 54\%, read 40\% \\
    C2 & 2,950 (12.8\%) & 60.39 & 15,039 & 1.09 & 0.50 & balanced + growth 12\% \\
    C3 & 8,083 (35.0\%) &  3.66 &    683 & 1.06 & 0.52 & comp 41\%, read 31\%, write 11\% \\
    C4 & 3,031 (13.1\%) &  3.74 &    144 & 0.90 & 0.49 & comp 69\%, read 8\% \\
    C5 & 3,151 (13.6\%) &  6.95 &  2,335 & 1.06 & 0.51 & growth 37\%, comp 28\% \\
    \bottomrule
  \end{tabular}
\end{table}

C1 has high transaction frequency, high-fee fraction, and transactions are dominated by computation and storage reads. C4 is the computation-dominant cluster. C5 represents wallets with significant storage growth.

Adversarial-activity validation for both L2 analyses is reported in Appendix~\ref{app:clust-validation}.

\section{Results: \eth L1}
\label{sec:l1}

\subsection{First Stage}

The instrument is strongly relevant on L1.
The overall first-stage coefficient is $\hat{\pi} = 0.480$. Although the partial $F$-statistics are inflated by the large sample size, the first-stage coefficient confirms a genuinely strong first stage, reflecting the strong autocorrelation in the daily L1 base fee.
The first-stage partial $R^2$ is $0.252$, confirming that the lagged base fee explains
a substantial share of current-fee variation even after partialling out the fixed effects.

\subsection{IV Estimates}

Table~\ref{tab:l1-iv} reports 2SLS estimates for the pooled sample and each
of the six clusters in Section~\ref{sec:l1_cluster_profiles}.
The pooled elasticity is $-0.0059^{***}$ (SE~$= 0.0005$): a $10\%$ increase
in the base fee reduces total L1 gas demand by approximately $0.06\%$.
L1 demand is near-inelastic. The aggregate near-inelasticity reflects the large share of low-intensity (C2, 26.1\%) and always-on (C3, 9.3\%) wallets whose consumption does not vary with the fee.
Given the size of the panel, statistical significance is easily attained; the economically relevant takeaway is that aggregate L1 gas demand is, for practical purposes, insensitive to fee changes.
The two most elastic clusters are C0 (34.0\%),
with $\hat{\beta} = -0.0236^{***}$ (SE~$= 0.0015$), and C5 (19.5\%), with $\hat{\beta} = -0.0212^{***}$ (SE~$= 0.0022$).
Wallets in C3 (9.3\%) have the highest first-stage $F$
($\hat{\pi} = 0.689$, $F = \num{15170218}$) yet the elasticity is near-zero
and statistically insignificant ($-0.0010$, $p = 0.17$). These are wallets
that do not respond to fee signals.

On L1, the OLS and IV estimates are close in magnitude: daily fixed effects already absorb most of the between-day congestion that drives the bias, so instrumenting moves the pooled estimate only slightly and the sign of the adjustment varies across clusters. The correction is larger on L2 (Section~\ref{sec:l2-total}), where hourly variation leaves more residual endogeneity.

\begin{table}[t]
  \centering
  \caption{\eth L1 IV elasticity estimates by cluster. Standard errors
    clustered at the wallet level.
    ${}^{**}p<0.01$, ${}^{***}p<0.001$; n.s.\ not significant at $5\%$.}
  \label{tab:l1-iv}
  \small
  \begin{tabular}{lrrrrr}
    \toprule
    Group & Wallets (\%) & $\hat{\beta}_{\text{OLS}}$
      & $\hat{\beta}_{\text{IV}}$ & SE & Partial $F$ \\
    \midrule
    Pooled            & \num{44449} (100\%)  & $-0.0074$ & $-0.0059^{***}$ & 0.0005 & \num{11873625} \\
    C0 (hi-fee act.)  & \num{15135} (34.0\%) & $-0.0110$ & $-0.0236^{***}$ & 0.0015 & \num{616714}   \\
    C1 (bursty)       &  \num{3609} (8.1\%)  & $+0.0011$ & $+0.0071$ n.s.  & 0.0051 & \num{111703}   \\
    C2 (low-intens.)  & \num{11625} (26.1\%) & $-0.0015$ & $-0.0016^{**}$  & 0.0006 & \num{1147259}  \\
    C3 (always-on)    &  \num{4123} (9.3\%)  & $-0.0022$ & $-0.0010$ n.s.  & 0.0007 & \num{15170218} \\
    C4 (variable)     &  \num{1280} (2.9\%)  & $-0.0116$ & $-0.0066$ n.s.  & 0.0036 & \num{298077}   \\
    C5 (fee-avoid.)   &  \num{8685} (19.5\%) & $-0.0308$ & $-0.0212^{***}$ & 0.0022 & \num{592378}   \\
    \bottomrule
  \end{tabular}
\end{table}

The negative slope is visible after removing wallet and day fixed effects,
confirming a downward-sloping demand relation in the partialled-out data. See full version of the paper.

\section{Results: Arbitrum L2 Aggregate Demand}
\label{sec:l2-total}

\subsection{First Stage}

The \arb total-gas panel uses hourly time fixed effects.
The pooled first-stage coefficient is $\hat{\pi} = 0.052$, significantly
lower than on \eth. This reflects the frequency with which \arb's base fee rests at its lower bound.
Because the partial $F$-statistic scales with sample size, its large value reflects
the size of the panel rather than instrument strength; the first-stage partial $R^2$
is only $0.007$, confirming that the lagged fee explains a negligible share of
current-fee variation on L2 after absorbing the fixed effects. The L2 estimates are
therefore identified from limited fee movement.

The pooled partial $F$-statistic is \num{118405}, well above the
commonly used threshold.
Per-cluster $F$-statistics range from \num{1361} (C0, high-volume but
sparsely active per hour) to \num{113875} (C1, long-running), confirming
the relevance property of the instrument.

\subsection{IV Estimates}

Table~\ref{tab:l2-total-iv} reports IV estimates for the \arb aggregate demand
model. The pooled elasticity is $-0.0359^{**}$ (SE~$= 0.0141$), six
times larger in magnitude than the \eth estimate.
A $10\%$ fee increase on \arb reduces total gas demand by an estimated
$0.36\%$, less elastic in absolute terms but measurably more responsive
than \eth.

The most credible cluster-level estimate is C1 (long-running high-gas wallets,
14.2\%), with $\hat{\beta} = -0.0782^{**}$ (SE~$= 0.0268$, $F = \num{113875}$).
The median number of active hours is highest in this cluster, giving the instrument strong identifying power.
The IV estimate for C2 (25.2\%) is near-zero and statistically insignificant
($-0.0057$, $p = 0.88$). Wallets in this cluster are occasional, low-frequency users that
do not adjust gas usage in response to fee variation.

We exclude C5 and C0 from causal interpretation.
C5 (28.0\%, median high-fee fraction~$= 0.90$)
yields a positive estimate ($+0.0314$, $p = 0.21$), consistent with exclusion
restriction failure: these wallets time activity to congestion periods, so
the lagged fee captures position during the previous congestion period.
C0 (11.3\%) yields $\hat{\beta} =
-1.163$, an anomalously large magnitude that is driven by very low active hours and is not identified.

\begin{table}[t]
  \centering
  \caption{\arb L2 aggregate IV elasticity estimates by cluster. Standard
    errors clustered at the wallet level.
    ${}^{**}p<0.01$, ${}^{***}p<0.001$; n.s.\ not significant at $5\%$;
    n.i.\ not identified.
    $\dagger$~excluded from causal interpretation.}
  \label{tab:l2-total-iv}
  \small
  \begin{tabular}{lrrrrr}
    \toprule
    Group & Wallets (\%) & $\hat{\beta}_{\text{OLS}}$
      & $\hat{\beta}_{\text{IV}}$ & SE & Partial $F$ \\
    \midrule
    Pooled                        & \num{23119} (100\%)  & $-0.0161$ & $-0.0359^{**}$  & 0.0141 & \num{118405} \\
    C0 (hi-vol.)$^{\dagger}$      &  \num{2606} (11.3\%) & $-0.0440$ & $-1.163$ n.i.   & 0.1795 & \num{1361}   \\
    C1 (long-run.)                &  \num{3279} (14.2\%) & $-0.0285$ & $-0.0782^{**}$  & 0.0268 & \num{113875} \\
    C2 (low-intens.)              &  \num{5818} (25.2\%) & $-0.0131$ & $-0.0057$ n.s.  & 0.0374 & \num{4662}   \\
    C3 (fee-avoid.)               &  \num{4253} (18.4\%) & $-0.0311$ & $+0.0792$ n.s.  & 0.0574 & \num{2709}   \\
    C4 (variable)                 &    \num{697} (3.0\%) & $-0.0306$ & $-0.1272$ n.s.  & 0.1687 & \num{2749}   \\
    C5 (cong.-act.)$^{\dagger}$   &  \num{6466} (28.0\%) & $-0.0102$ & $+0.0314$ n.s.  & 0.0249 & \num{7163}   \\
    \bottomrule
  \end{tabular}
\end{table}

\rev{The pooled estimate includes the C0 and C5 wallets whose instruments are
potentially invalid (Section~\ref{sec:iv}). Since C5 contributes a near-zero
positive estimate and C0 an anomalously large negative one, the pooled $-0.036$
is an average over heterogeneous reliability; the valid-cluster estimates in
Table~\ref{tab:l2-total-iv} provide the within-cluster causal figures.}

\rev{A natural alternative identification strategy would exploit the
0.01~$\to$~0.02~gwei base fee floor increase mid-sample as a quasi-experimental
discontinuity. This change, however, binds only for transactions priced at or
near the floor --- primarily spam and dust activity --- while the 500-transaction
wallets in our sample operate well above the floor throughout. The floor increase
therefore does not induce a sharp discontinuity in the fee faced by our study
population and cannot serve as an instrument for this sample.}

The two-way demeaned binscatter confirms a negative demand slope after removing wallet and hour fixed effects and log transaction count (see full version).

\section{Results: Arbitrum L2 Per-Resource Demand}
\label{sec:l2-multidim}

\subsection{Per-Dimension Pooled Estimates}

Table~\ref{tab:multidim-pooled} reports IV elasticity estimates for each of
the seven \arb resource dimensions using the pooled sample.
The estimates span a wide range across resource dimensions.
These are intensive-margin elasticities, estimated on wallet-hours with positive
consumption of the given resource; they describe the response of wallets that
continue to use a resource, not the economy-wide change in its total consumption.
Computation gas, the dominant component of most transactions, is modestly
elastic at $-0.0274^{*}$ (SE~$= 0.0123$).
Storage growth and refunds are the most price-sensitive: $-0.1510^{***}$
(SE~$= 0.0446$) and $-0.2688^{***}$ (SE~$= 0.0414$) respectively.
L2 calldata shows a significant negative elasticity of $-0.0575^{*}$
(SE~$= 0.0228$).
History growth is statistically insignificant from zero ($p = 0.72$).

\begin{table}[t]
  \centering
  \caption{L2 per-resource IV elasticity estimates, pooled sample.
    Intensive margin (wallet-hours with $\mathrm{gas}_d > 0$).
    ${}^{*}p<0.05$, ${}^{***}p<0.001$; n.s.\ not significant at $5\%$.}
  \label{tab:multidim-pooled}
  \small
  \begin{tabular}{lrrrrr}
    \toprule
    Dimension & Wallets & $\hat{\beta}_{\text{OLS}}$
      & $\hat{\beta}_{\text{IV}}$ & SE & Partial $F$ \\
    \midrule
    computation        & \num{23119} & $-0.0026$ & $-0.0274^{*}$   & 0.0123 & \num{118405} \\
    historyGrowth      & \num{20052} & $-0.0109$ & $-0.0109$ n.s.  & 0.0304 & \num{46410}  \\
    storageAccessRead  & \num{22652} & $+0.0166$ & $+0.0344^{*}$   & 0.0153 & \num{118185} \\
    storageAccessWrite & \num{20883} & $+0.0040$ & $+0.1186^{***}$ & 0.0274 & \num{47843}  \\
    storageGrowth      & \num{18990} & $-0.0388$ & $-0.1510^{***}$ & 0.0446 & \num{26104}  \\
    l2Calldata         & \num{20799} & $-0.0125$ & $-0.0575^{*}$   & 0.0228 & \num{71727}  \\
    refund             & \num{17770} & $-0.0232$ & $-0.2688^{***}$ & 0.0414 & \num{30666}  \\
    \bottomrule
  \end{tabular}
\end{table}

\subsection{Positive Elasticities}

The positive IV estimates for storage access reads ($+0.034^{*}$) and writes
($+0.119^{***}$) do not represent a genuine demand response.
When fees rise, wallets with lower-necessity operations exit the sample. This leaves a selected set of wallets whose operations require storage access.
In such periods, the conditional average storage access rises even as total volume falls, producing a spurious positive elasticity.
Because this selection operates whenever fees rise, it biases every per-resource estimate in the positive direction. The dimensions that remain significantly negative, storage growth and refunds, are therefore conservative: the true intensive-margin response is at least as large as the estimates reported here. The small or positive estimates for computation, reads, and writes should be read as responses among continuing wallets rather than economy-wide demand elasticities.

The share of active wallet-hours using storage reads is effectively constant across
base-fee deciles (0.951--0.956), confirming that read-intensive wallets are the last
to exit when fees rise and dominate the remaining sample at high fees.
Storage writes show a clearer decline (0.835 at the lowest fee decile to 0.773 at
the highest), consistent with write-light wallets exiting while write-intensive
operations persist.
By contrast, the dimensions with the most negative IV estimates, storage growth and
refunds, show no systematic selection pattern across deciles, supporting their
interpretation as genuine demand responses rather than selection artifacts.

\subsection{Most Elastic Resources: Storage Growth and Refunds}

Storage growth ($\hat{\beta} = -0.1510^{***}$) and refunds ($\hat{\beta} =
-0.2688^{***}$) are the most price-sensitive resource dimensions in the pooled estimate.
Both effects are concentrated in C3 of the per-resource clustering (35.0\% of the sample): $-0.3709^{***}$ for storage growth
(SE~$= 0.0549$) and $-0.4157^{***}$ for refunds (SE~$= 0.0514$).

The wallets in this cluster have routine operations for which creating new storage slots and storage cleanup are not time sensitive. When fees rise, they defer these operations, which lead to a measurable reduction in usage of these resources.

\subsection{Cluster-Level Per-Dimension Results}

Table~\ref{tab:multidim-clusters} presents cluster-level estimates for the two
clusters with the most interpretable IV results.
C2 (12.8\%) shows large negative elasticities across
nearly all dimensions except storage growth and refunds, confirming price
elasticity in this segment.
C3 (35.0\%) is near-zero across most dimensions but strongly
negative for storage growth and refunds, consistent with the pattern observed in the pooled sample.

\begin{figure}[t]
  \centering
  \includegraphics[width=0.88\linewidth]{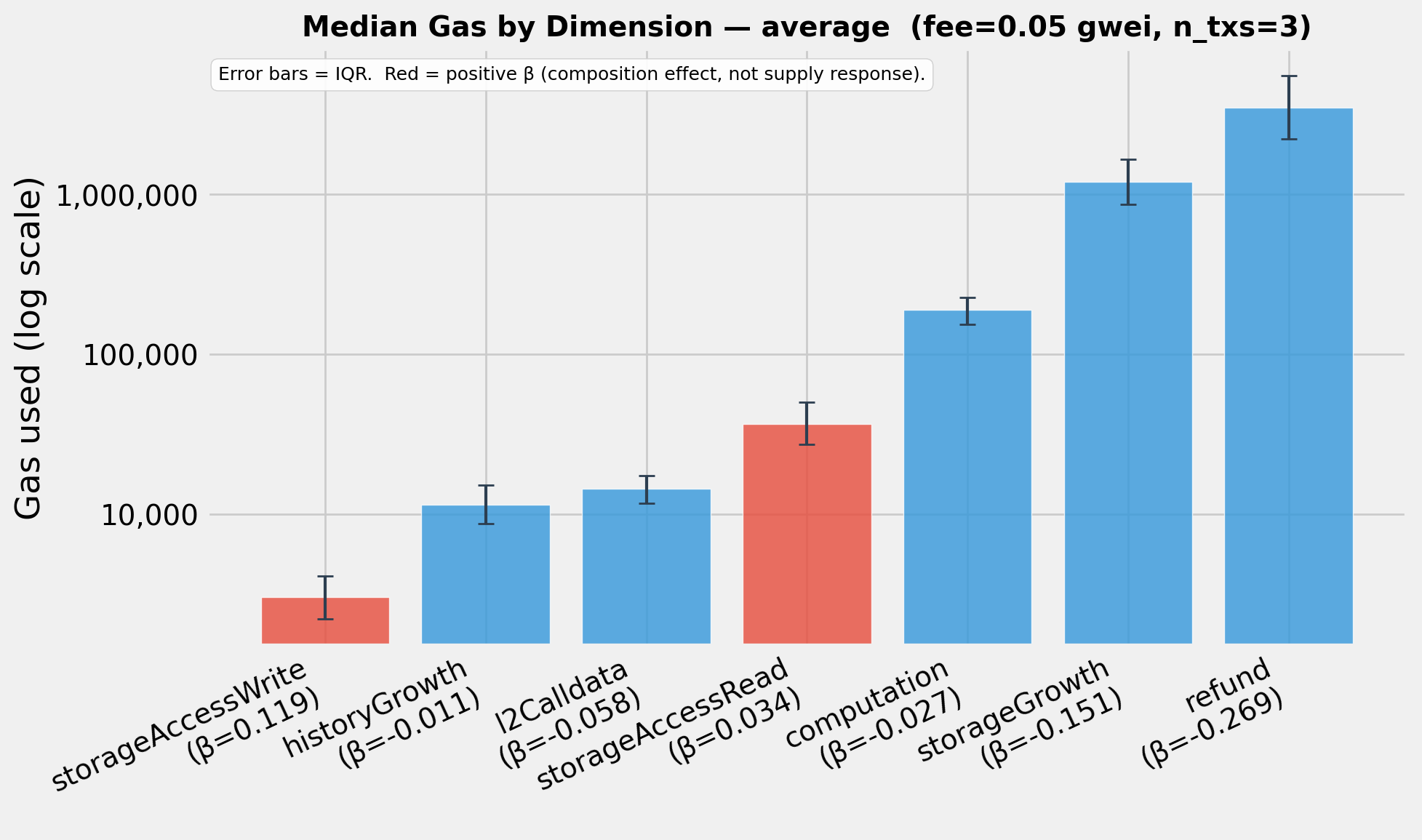}
  \caption{Median gas usage by resource dimension for the pooled \arb L2
    model (\texttt{average} regime, fee~$= 0.05$~gwei).
    Error bars show the inter-quartile range.
    Blue bars have negative $\hat{\beta}$ (demand response);
    red bars have positive $\hat{\beta}$.}
  \label{fig:dim-comparison}
\end{figure}

Figure~\ref{fig:dim-comparison} summarizes median gas usage and the pooled
IV elasticity across dimensions.

\begin{table}[t]
  \centering
  \caption{L2 per-resource IV $\hat{\beta}$ for key clusters.
    ${}^{*}p<0.05$, ${}^{***}p<0.001$; n.s.\ not significant at $5\%$.}
  \label{tab:multidim-clusters}
  \small
  \begin{tabular}{lrr}
    \toprule
    Dimension & C2 high-volume (12.8\%) & C3 general DeFi (35.0\%) \\
    \midrule
    computation        & $-0.1590^{***}$ & $-0.0232$ n.s.  \\
    historyGrowth      & $-0.6859^{***}$ & $-0.0354$ n.s.  \\
    storageAccessRead  & $-0.3189^{***}$ & $-0.0274$ n.s.  \\
    storageAccessWrite & $-0.3207^{***}$ & $+0.0069$ n.s.  \\
    storageGrowth      & $+0.1274$ n.s.  & $-0.3709^{***}$ \\
    l2Calldata         & $-0.3560^{***}$ & $+0.0107$ n.s.  \\
    refund             & $+0.0781$ n.s.  & $-0.4157^{***}$ \\
    \bottomrule
  \end{tabular}
\end{table}

\section{Downstream applications}
\label{sec:applications}

\begin{figure}[t]
  \centering
  \includegraphics[width=0.78\linewidth]{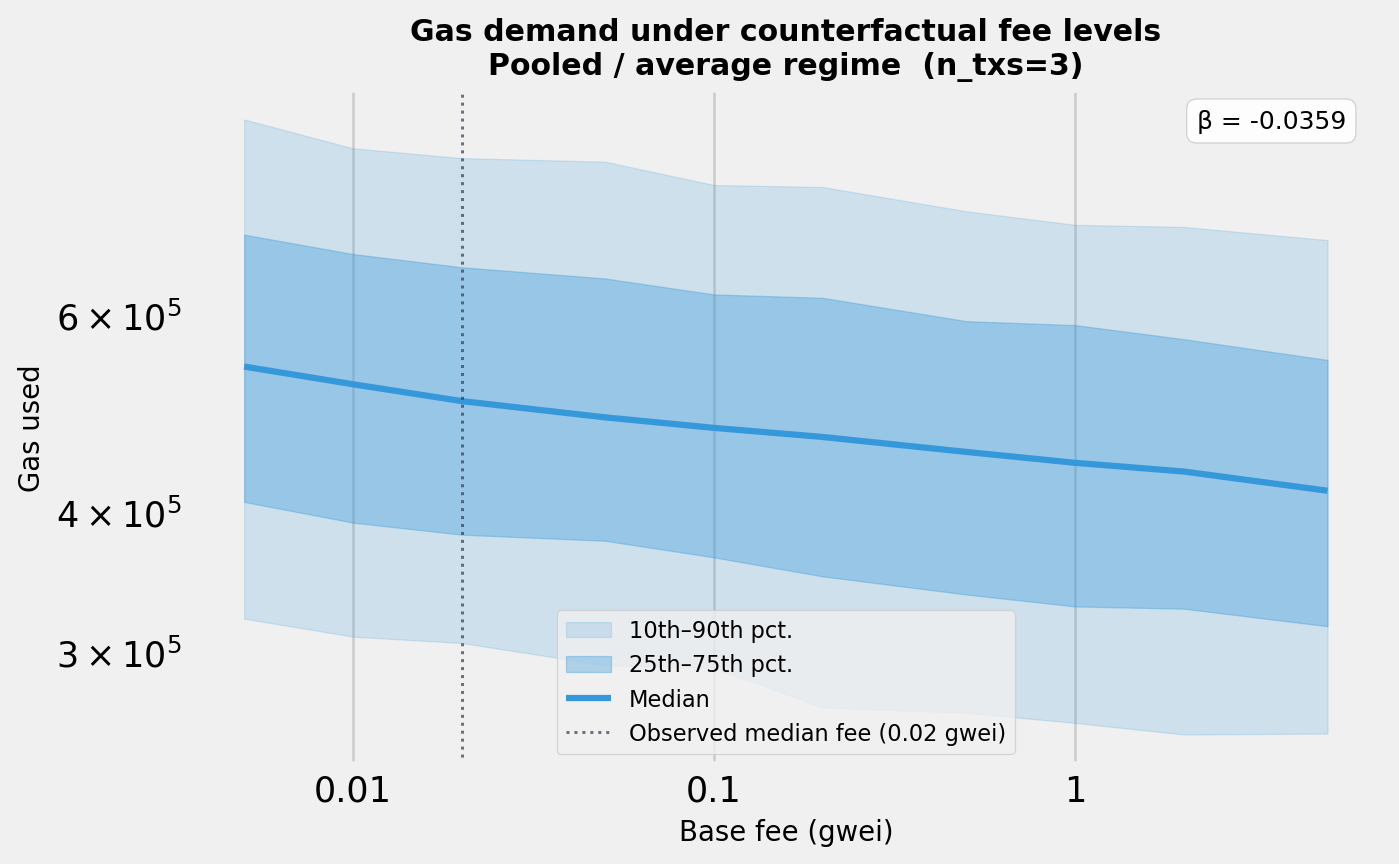}
  \caption{Simulated gas demand under counterfactual base fees, \texttt{average} regime.
    Bold line: median. Inner band: 25th--75th percentile. Outer band: 10th--90th percentile.
    Dotted vertical line: observed median base fee ($0.02$~gwei).}
  \label{fig:fee-sweep}
\end{figure}

The IV estimates from Section~\ref{sec:iv} include the causal parameters of a demand model that can be used to simulate gas usage by the wallets in various gas fee regimes. In this section, we describe how the full demand model is used to generate synthetic gas demand distributions under counterfactual fee regimes. Such simulations help evaluate the performance of fee mechanism designs beyond the observed range. We present the construction of the demand model here.
Figure~\ref{fig:fee-sweep} illustrates the model applied to a fee sweep from
$0.005$ to $5$~gwei for the \texttt{average} regime
($\hat\alpha \in [P_{40}, P_{60}]$, $\hat\gamma \in [P_{40}, P_{60}]$,
$n_{\text{txs}}=3$).
The downward-sloping median response reflects the pooled L2 elasticity
($\hat\beta = -0.036$) across the full fee range; the interquartile and
10th--90th percentile envelopes shift in parallel, confirming that the elasticity
acts multiplicatively on the gas distribution rather than compressing or expanding it.
A systematic evaluation of specific mechanism parameters is left to future work.

\subsection{Recovering Fixed Effects}
\label{sec:recover_fe}

The 2SLS method described in Section~\ref{sec:2sls} estimates only $\hat{\beta}$ and $\hat{\phi}$ of the structural
demand equation~\eqref{eq:structural}. The wallet and time fixed effects are recovered by computing the residuals of the outcome and applying alternating projections. The residuals are then fitted with a Student-$t$ distribution to accommodate heavy tails that are commonly observed in gas usage. The alternating projections procedure and the sampling algorithm used to generate counterfactual gas demand distributions are detailed in the full version.

\section{Discussion}
\label{sec:discussion}

\subsection{Why L2 Is More Elastic Than L1}

Several factors may contribute to the higher measured elasticity on L2. The two chains attract different wallet populations: L2 users have opted into a lower-cost environment and may be more price-sensitive by selection, whereas L1 wallets include a higher share of protocol-level operations that are insensitive to fee changes. Arbitrum's base fee floor and gas ladder also create a distinct incentive environment compared to Ethereum. Finally, the difference in time granularity between the two estimates, daily fixed effects on L1 versus hourly on L2, may mechanically affect the measured elasticity in ways that are difficult to disentangle from a genuine behavioral difference.

One particular difference is the relative absence of a large share of reverted transactions on L1, unlike L2s, where the sequencer's mempool is kept private and arbitrageurs try to probabilistically backrun state changes. In times of high asset price volatility, which results in higher on-chain fees, such transactions dominate on-chain gas consumption. On L1, the task of filtering reverted transactions is delegated to the block builders.

\subsection{Implications for Fee Mechanism Design}

Ethereum and Arbitrum both use a variant of the same congestion management fee mechanism, namely, converting all resource usage into a single gas unit and pricing the gas according to its own EIP-1559 parameters.
Note that not all resources are the same. Namely, state growth is a long-term issue for the chain: if the state grows too much, accessing and writing to state memory become slower, and it may no longer fit on disk. Other resources have more immediate constraints. Storage capacity not allocated at a given time can be allocated later, whereas computation and state-access resources not consumed in a given period cannot be recovered. These distinctions call for at least different ratios between limit and target per resource, instead of a constant ratio imposed by converting all resources to a single gas unit, to better utilize the chain. Fully-fledged multi-dimensional resource pricing schemes have been proposed and analyzed by a number of industry and academic works. Our findings on multi-dimensional resource elasticities can help inform optimal parameter ranges for each resource, given EIP-1559-type mechanisms are applied to each resource independently. As an illustration, the comparatively elastic refund and storage-growth dimensions (pooled $-0.27$ and $-0.15$) imply that pricing these resources more aggressively would meaningfully curb their use, whereas repricing computation (near-zero elasticity) would mainly raise revenue without changing usage. A multi-dimensional mechanism could therefore apply tighter independent pricing to the elastic, state-affecting resources, where price has traction.


\subsection{Limitations}

The IV estimator identifies a local average treatment effect: the demand response
of wallets whose current fee is shifted by the lagged fee. Always-on wallets (C3 on
L1) and congestion-timing wallets (C5 on L2) are excluded from causal
interpretation; the pooled estimates do not characterize these populations. The L2
sample covers seven months (October 2025 through April 2026) and reflects the fee
regime in effect during this window; as Arbitrum's base fee floor and gas ladder
parameters evolve, the measured elasticities may shift. The 500-transaction minimum
threshold ensures sufficient within-wallet variation for fixed-effects estimation
but excludes low-frequency users, and the estimated elasticities may not generalize
to that population. Finally, the per-resource decomposition is feasible for Arbitrum
because the Nitro node tracks per-category gas consumption; no equivalent data
exists for Ethereum L1.

\rev{The methodology applies to any chain with a congestion-responsive fee
adjustment rule (an EIP-1559 variant or equivalent), sufficient wallet-level
panel data, and temporal fee variation above the protocol floor. For chains with
fixed-price auctions or purely flat fees, the lagged-fee instrument loses
relevance and a different identification strategy would be required. Chains where
the base fee spends most of its time at the floor (as L2 does during
low-congestion periods) should use the partial $R^2$ alongside the partial $F$
to assess instrument strength, as shown here for \arb.}

\section{Related Work}
\label{sec:related}

This work draws on, and contributes to, four research areas: theoretical and empirical study of transaction fee mechanisms, multi-dimensional resource pricing, L2 fee economics, and the empirical estimation of demand elasticity under congestion pricing.

\subsection{Transaction Fee Mechanism Design}
Roughgarden's analysis of \onefivefivenine~\cite{roughgarden2021eip1559,roughgarden2021mechanism} establishes the design objectives of the protocol's base-fee adjustment rule and characterizes its incentive properties. Ferreira et al.~\cite{ferreira2021dynamic} develop a class of dynamic posted-price mechanisms for blockspace, and Chung and Shi~\cite{chung2023} provide a unified treatment of impossibility results for off-chain incentive-compatible transaction fee mechanisms. Leonardos et al.~\cite{leonardos2021} give a control-theoretic characterization of base-fee dynamics under stochastic demand. The original proposal~\cite{buterin2019eip1559} motivates the mechanism in terms of wallet-level fee predictability, the property that underpins the present empirical strategy: predictable fees make the lagged base fee a relevant instrument for the current fee.

\subsection{Empirical Studies of Gas Fees and Demand}
Reijsbergen et al.~\cite{reijsbergen2021} examine the first month after \onefivefivenine deployment and document a substantial reduction in fee volatility with limited change in median fees. Liu et al.~\cite{liu2022} extend this horizon and quantify the effect of the mechanism on waiting times and consensus security. Pierro and Rocha~\cite{pierro2019} provide an earlier empirical decomposition of gas-fee covariates including pending transaction count and exogenous costs. The closest predecessor to the present elasticity estimate is Donmez and Karaivanov~\cite{donmez2022}, who use block-level data from the pre-\onefivefivenine first-price regime and a queueing-theoretic specification to identify a strongly nonlinear utilization-price relationship, with sharp fee response above $90\%$ block fullness. Their analysis predates \onefivefivenine, does not exploit wallet-level variation, and does not separate intensive from extensive margins; the present work addresses each of these in the post-\onefivefivenine regime and additionally provides per-resource estimates for an L2.
Silva~\cite{silva2026elasticity} studies multi-resource demand elasticities on Ethereum, complementing our data collection; the focus is on the effects on per-resource demand as the Ethereum gas target increases.

\subsection{Multi-Dimensional Resource Pricing}

Diamandis et al.~\cite{diamandis2023multidim} introduce multi-dimensional EIP-1559-style pricing for non-fungible blockchain resources and characterize the equilibrium fee schedule. Angeris et al.~\cite{angeris2025optimal} prove that multi-dimensional fees are essentially optimal in a broad class of demand models. Lavee et al.~\cite{multi-dimensional} and Kiayias et al.~\cite{one_vs_multi} compare the throughput and convergence properties of multi-dimensional versus single-dimensional pricing: Lavee et al.\ study stationary throughput and find that the ranking depends on demand functions, while Kiayias et al.\ take a dynamic approach and argue that a single-dimensional mechanism may reach equilibrium prices faster. Crapis et al.~\cite{optimal_dynamic_fees} study optimal design of a multi-dimensional pricing mechanism when resource usages are pairwise correlated.

\subsection{Rollup Economics and MEV on L1/L2}

Monnot~\cite{monnot2022rollup} and Crapis~\cite{crapis2023rollup} provide foundational economic analyses of rollup fee markets. Crapis et al.~\cite{l2_posting_on_l1} study optimal strategies of L2 data posting on L1 and effects on end users.

Solmaz et al.~\cite{optimistic_mev} empirically find that a large percentage of L2 gas usage is due to optimistic MEV, which is not observed on L1 since the block builders can and do filter out reverted transactions. This phenomenon is theoretically studied in Mazorra et al.~\cite{timing_games} and Wang et al.~\cite{blockspace_pressure}, where a similar conclusion is made: given fixed gas costs, probabilistic backrunning searchers will spend all expected value in reverted transactions (sometimes classified as spam).

\subsection{Wallet Identification and Behavioral Clustering}
In an account-based blockchain such as \eth, identifying wallet type is a prerequisite for any heterogeneity analysis. Victor~\cite{victor2020clustering} introduces the standard address-clustering heuristics for \eth. B\'eres et al.~\cite{beres2021profiling} extend this with timing- and graph-based features for finer profiling of addresses. Hu et al.~\cite{hu2024bots} provide recent feature-engineering work specifically aimed at separating financial bots and MEV searchers from human users. The clustering procedure in Section~\ref{sec:clustering} adopts features in this tradition (transaction frequency, gas variance, and high-fee timing) and augments them with per-resource consumption fractions for the multi-dimensional analysis.

\subsection{Demand Elasticity under Congestion Pricing}
The methodological framework used to estimate elasticity in this work has direct precedence in the electricity pricing literature. Borenstein~\cite{borenstein2005realtime} establishes that even small demand elasticities (around $-0.025$, a range comparable to the L1 elasticity estimate) translate to substantial welfare gains under dynamic pricing. Fabra et al.~\cite{fabra2021elasticity} use household-level panel data and a renewable-supply instrument to estimate elasticity under Spain's real-time pricing tariff, providing the closest methodological analogue to our wallet-level panel analysis. Tiedemann et al.~\cite{tiedemann2024iv} show that standard IV is inconsistent when high-frequency time series exhibit autocorrelation in both price and demand, informing the within-period exclusion concerns discussed in Section~\ref{sec:iv}. The foundations of IV analysis are described in~\cite{angristkrueger1991,staiger1997,wooldridge2010,angrist2009}.
\section{Conclusion}
\label{sec:conclusion}

We estimate the IV elasticities of gas demand for Ethereum
L1 and Arbitrum L2.
Gas demand is inelastic on both networks: pooled IV estimates of $-0.006$ (L1)
and $-0.036$ (L2) imply that a $10\%$ fee increase reduces total demand by at
most $0.4\%$.
Within L2, per-resource decomposition reveals that the aggregate inelasticity
masks substantial resource-level heterogeneity: refunds and new storage slot
creation respond meaningfully to price ($-0.27$ and $-0.15$ respectively),
while computation is near-inelastic ($-0.03$).

\bibliography{refs}

\appendix

\section{Full Per-Cluster IV Tables}
\label{app:tables}

Full per-cluster IV estimates for L1 and L2 aggregate demand appear in
Tables~\ref{tab:l1-iv} and~\ref{tab:l2-total-iv} in the main body.
Tables~\ref{tab:multidim-full-c012} and~\ref{tab:multidim-full-c345} below
report IV elasticity estimates for all six clusters across all seven
\arb resource dimensions.

\begin{table}[t]
  \centering
  \caption{L2 per-resource IV $\hat{\beta}_{\text{IV}}$ (SE) for clusters C0, C1, C2.
    ${}^{*}p<0.05$, ${}^{**}p<0.01$, ${}^{***}p<0.001$; n.s.\ not significant at $5\%$.}
  \label{tab:multidim-full-c012}
  \small
  \begin{tabular}{lrrr}
    \toprule
    Dimension
      & C0 hi-vol.\ burst (3.3\%)
      & C1 cong.-active (22.2\%)
      & C2 high-volume (12.8\%) \\
    \midrule
    computation        & $-0.022$ (0.131) n.s.  & $-0.026$ (0.015) n.s.  & $-0.159^{***}$ (0.026) \\
    historyGrowth      & $-0.314$ (0.169) n.s.  & $+1.105$ (0.820) n.s.  & $-0.686^{***}$ (0.092) \\
    storageAccessRead  & $-0.086$ (0.136) n.s.  & $+0.280^{***}$ (0.033) & $-0.319^{***}$ (0.033) \\
    storageAccessWrite & $+0.104$ (0.128) n.s.  & $+3.534^{***}$ (0.998) & $-0.321^{***}$ (0.077) \\
    storageGrowth      & $-0.429$ (0.326) n.s.  & $-0.131$ (0.459) n.s.  & $+0.127$ (0.096) n.s.  \\
    l2Calldata         & $-0.295^{*}$ (0.134)   & $+0.906^{**}$ (0.288)  & $-0.356^{***}$ (0.033) \\
    refund             & $+0.215$ (0.233) n.s.  & $+0.637$ (1.211) n.s.  & $+0.078$ (0.094) n.s.  \\
    \bottomrule
  \end{tabular}
\end{table}

\begin{table}[t]
  \centering
  \caption{L2 per-resource IV $\hat{\beta}_{\text{IV}}$ (SE) for clusters C3, C4, C5.
    Significance as in Table~\ref{tab:multidim-full-c012}.}
  \label{tab:multidim-full-c345}
  \small
  \begin{tabular}{lrrr}
    \toprule
    Dimension
      & C3 general DeFi (35.0\%)
      & C4 variable gas (13.1\%)
      & C5 storage-growth (13.6\%) \\
    \midrule
    computation        & $-0.023$ (0.022) n.s.  & $+0.081^{*}$ (0.037)   & $+0.012$ (0.079) n.s.  \\
    historyGrowth      & $-0.035$ (0.035) n.s.  & $+0.477^{***}$ (0.106) & $+0.095$ (0.103) n.s.  \\
    storageAccessRead  & $-0.027$ (0.027) n.s.  & $+0.086$ (0.107) n.s.  & $+0.151$ (0.101) n.s.  \\
    storageAccessWrite & $+0.007$ (0.034) n.s.  & $+0.378^{*}$ (0.153)   & $+0.076$ (0.088) n.s.  \\
    storageGrowth      & $-0.371^{***}$ (0.055) & $+0.383$ (0.429) n.s.  & $+0.042$ (0.109) n.s.  \\
    l2Calldata         & $+0.011$ (0.029) n.s.  & $+0.301^{***}$ (0.085) & $-0.145$ (0.095) n.s.  \\
    refund             & $-0.416^{***}$ (0.051) & $+0.210$ (0.352) n.s.  & $+0.123$ (0.092) n.s.  \\
    \bottomrule
  \end{tabular}
\end{table}

\section{Iterative Demeaning and Simulation Details}
\label{app:demeaning}

\subsection{Alternating Projections}
\label{app:alternating}

Given 2SLS estimates $\hat{\beta}$ and $\hat{\phi}$, the residual log-gas is
\begin{equation}
  \label{eq:residual}
  r_{it} = \log g_{it} - \hat{\beta} \log p_{it} - \hat{\phi} \log n_{it},
\end{equation}
which satisfies $r_{it} \approx \alpha_i + \gamma_t + \varepsilon_{it}$.
The fixed effects are recovered by alternating projections, initializing
$\gamma_t = 0$ and iterating:
\begin{align}
  \alpha_i &\leftarrow \frac{1}{|T_i|}\sum_{t \in T_i} \bigl(r_{it} - \gamma_t\bigr), \\
  \gamma_t &\leftarrow \frac{1}{|I_t|}\sum_{i \in I_t} \bigl(r_{it} - \alpha_i\bigr),
\end{align}
where $T_i$ is the set of time periods in which wallet $i$ is observed and
$I_t$ is the set of wallets active in period $t$.

The procedure is repeated until convergence. In all cases, the procedure
converged (to a tolerance level of $10^{-8}$) within 500 iterations.
After convergence, $\gamma_t$ and $\alpha_i$ are recentred:
$\gamma_t \leftarrow \gamma_t - \bar{\gamma}$ and
$\alpha_i \leftarrow \alpha_i + \bar{\gamma}$, where
$\bar{\gamma} = \frac{1}{|T|}\sum_t \gamma_t$.

The remaining residuals
\begin{equation}
  \hat{\varepsilon}_{it} = r_{it} - \hat{\alpha}_i - \hat{\gamma}_t
\end{equation}
are then fitted by MLE to a Student-$t$ distribution with
location $\hat{\mu}$, scale $\hat{\sigma}$, and degrees of freedom $\hat{\nu}$.

\subsection{Sampling Procedure}
\label{app:sampling}

After recovering all parameters in Equation~\eqref{eq:structural}, a gas
demand sample for a base fee $p$ and transaction count $n$ is drawn as
follows. A \textit{regime} is first selected, defined by percentile windows
on the empirical distributions of $\hat{\alpha}_i$ and $\hat{\gamma}_t$.
Table~\ref{tab:regimes} lists the seven regimes used in this analysis.

\begin{table}[t]
  \centering
  \caption{Simulation regimes. Each regime is defined by percentile windows
    $[a_\ell, a_u]$ on $\hat{\alpha}_i$ (wallet type) and $[g_\ell, g_u]$ on
    $\hat{\gamma}_t$ (network condition).}
  \label{tab:regimes}
  \small
  \begin{tabular}{lcccc}
    \toprule
    Regime & $a_\ell$ & $a_u$ & $g_\ell$ & $g_u$ \\
    \midrule
    average                  & 40 & 60 & 40 &  60 \\
    heavy\_wallet            & 80 & 100 & 40 &  60 \\
    light\_wallet            &  0 &  20 & 40 &  60 \\
    peak\_fee                & 40 &  60 & 80 & 100 \\
    quiet                    & 40 &  60 &  0 &  20 \\
    heavy\_wallet\_peak\_fee & 80 & 100 & 80 & 100 \\
    light\_wallet\_quiet     &  0 &  20 &  0 &  20 \\
    \bottomrule
  \end{tabular}
\end{table}

A gas draw for fee $p$, transaction count $n$, and $n_{\text{samples}}$
replications is produced by:
\begin{enumerate}
  \item Draw $\hat{\alpha}^{(k)} \sim \mathrm{Uniform}(\mathcal{A})$ and
    $\hat{\gamma}^{(k)} \sim \mathrm{Uniform}(\mathcal{G})$,\;
    $k = 1, \ldots, n_{\text{samples}}$, where $\mathcal{A}$ and $\mathcal{G}$
    are the empirical distributions of $\hat{\alpha}_i$ and $\hat{\gamma}_t$
    restricted to the selected regime.
  \item Draw $\hat{\varepsilon}^{(k)} \sim t(\hat{\nu},\, \hat{\mu},\, \hat{\sigma})$.
  \item Compute
    $\log \tilde{g}^{(k)} = \hat{\alpha}^{(k)} + \hat{\gamma}^{(k)}
      + \hat{\beta} \log p + \hat{\phi} \log n + \hat{\varepsilon}^{(k)}$.
  \item Return $\tilde{g}^{(k)} = \exp\!\bigl(\log \tilde{g}^{(k)}\bigr)$.
\end{enumerate}

\section{Exclusion Restriction Discussion}
\label{app:exclusion}

The instrument $z_{it} = \log p_{i,t-1}$ is the wallet's own average base fee during its previous active period. Two properties support its exogeneity with respect to the current demand shock $u_{it}$. First, the base fee is a global quantity set by aggregate block utilization; an individual wallet's transactions are a negligible fraction of total block gas, so the lagged fee is a market-level price realized before period $t$ rather than a function of the wallet's own demand. Second, it is predetermined: being a function of utilization in periods strictly before $t$, it is fixed before the wallet makes its period-$t$ usage decision and cannot respond to $u_{it}$.

The exclusion restriction can still fail for wallets that systematically time their activity to congestion. For such a wallet the lagged fee marks the level of a prior congestion episode, and if congestion is serially correlated the lagged fee carries information about the current demand shock. The two-way fixed effects absorb the wallet's average timing and the period's average congestion but not their interaction. We mitigate this by excluding wallets with an anomalously high fraction of activity in above-median-fee periods (the high-fee-fraction cluster, Section~\ref{sec:iv}), for which the channel is most plausible; the retained wallets transact across fee levels in a manner not systematically aligned with congestion.

Table~\ref{tab:exclusion-timing} (main body, Section~\ref{sec:iv}) quantifies this timing
differential. The figure below shows the full distribution by cluster.

\begin{figure}[t]
  \centering
  \includegraphics[width=0.78\linewidth]{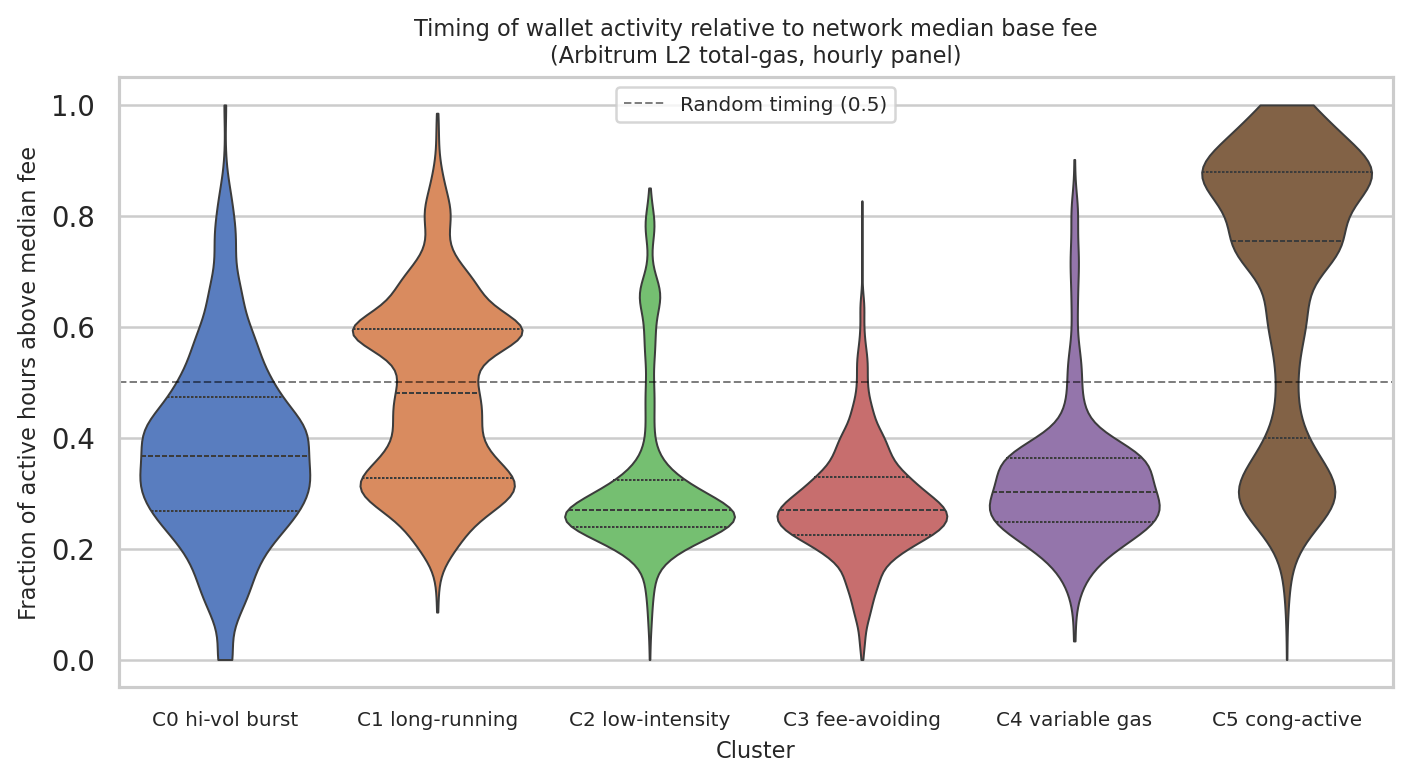}
  \caption{Distribution of the fraction of active hours in above-median-fee periods,
    by cluster. C5 is a clear outlier; the five retained clusters are near or below
    the random baseline of 0.50.}
  \label{fig:exclusion-timing}
\end{figure}

\section{Clustering Diagnostics}
\label{app:clustering}

Figures~\ref{fig:elbow-l1}, \ref{fig:elbow-l2}, and~\ref{fig:elbow-multidim} show
the within-cluster sum of squared distances (inertia) as a function of the number
of clusters $k$ for each of the three analyses: L1 total-gas, L2 total-gas, and L2
per-resource. In all three cases the inertia curve bends sharply at $k = 6$ before
flattening, indicating that additional clusters beyond six provide diminishing
reduction in inertia. The $k = 6$ choice is used for all analyses reported in the
main text.

\input{aft_paper/paper/elbow_pictures.tex}

\section{Cluster Validation}
\label{app:clust-validation}

We validate the clusters using a classification derived from on-chain activity.
A wallet is flagged as \adv if it meets either of two thresholds: (i) the wallet
is in the top 0.1 percentile on that day based on transaction count, or (ii) the
wallet's revert ratio on that day is $\geq 50\%$ with at least 5 transactions. The
latter signal is characteristic of wallets participating in competitive maximal
extractable value (MEV) strategies. A wallet is considered \adv if at least $50\%$
of its active days are flagged, and \textit{\adv-ever} if at least one day is flagged.

\subsubsection*{L2 total gas clusters}

C1 (long-running) is the most \adv cluster: $64.5\%$ of wallets were flagged at
some point, $34.2\%$ remain flagged currently, with a median revert ratio of
$2.9\%$. This is the only cluster with a non-zero median \adv-day fraction ($10.8\%$).
C5 (congestion-active) shows a high \adv-ever rate ($43.8\%$) but near-zero current
prevalence ($0.8\%$) and zero median \adv days, indicating wallets that were
previously active in competitive periods but have since reduced activity.
C2 (low-intensity, $8.3\%$ ever) and C3 (fee-avoiding, $6.8\%$ ever) are
predominantly non-\adv, matching their interpretation as retail and price-sensitive
wallets.

\begin{table}[t]
  \centering
  \caption{\Adv activity by cluster, \arb L2 total-gas analysis.
    \textit{\Adv-ever}: fraction flagged at any point.
    \textit{Current}: fraction flagged at classification time.
    Revert ratio and \adv days are medians over matched wallets.}
  \label{tab:adv-l2-total}
  \small
  \begin{tabular}{llrrrr}
    \toprule
    Cluster & Description & \makecell{\Adv-ever\\(\%)} & \makecell{Current\\(\%)}
      & \makecell{Med.\ revert\\(\%)} & \makecell{Med.\ \adv\\days (\%)} \\
    \midrule
    C0 & Hi-vol burst      & 35.2 & 16.5 & 2.09 &  0.0 \\
    C1 & Long-running      & 64.5 & 34.2 & 2.90 & 10.8 \\
    C2 & Low-intensity     &  8.3 &  0.3 & 0.00 &  0.0 \\
    C3 & Fee-avoiding      &  6.8 &  1.7 & 0.12 &  0.0 \\
    C4 & Variable gas      & 19.9 &  2.5 & 0.02 &  0.0 \\
    C5 & Congestion-active & 43.8 &  0.8 & 0.87 &  0.0 \\
    \bottomrule
  \end{tabular}
\end{table}

\begin{figure}[t]
  \centering
  \includegraphics[width=0.88\linewidth]{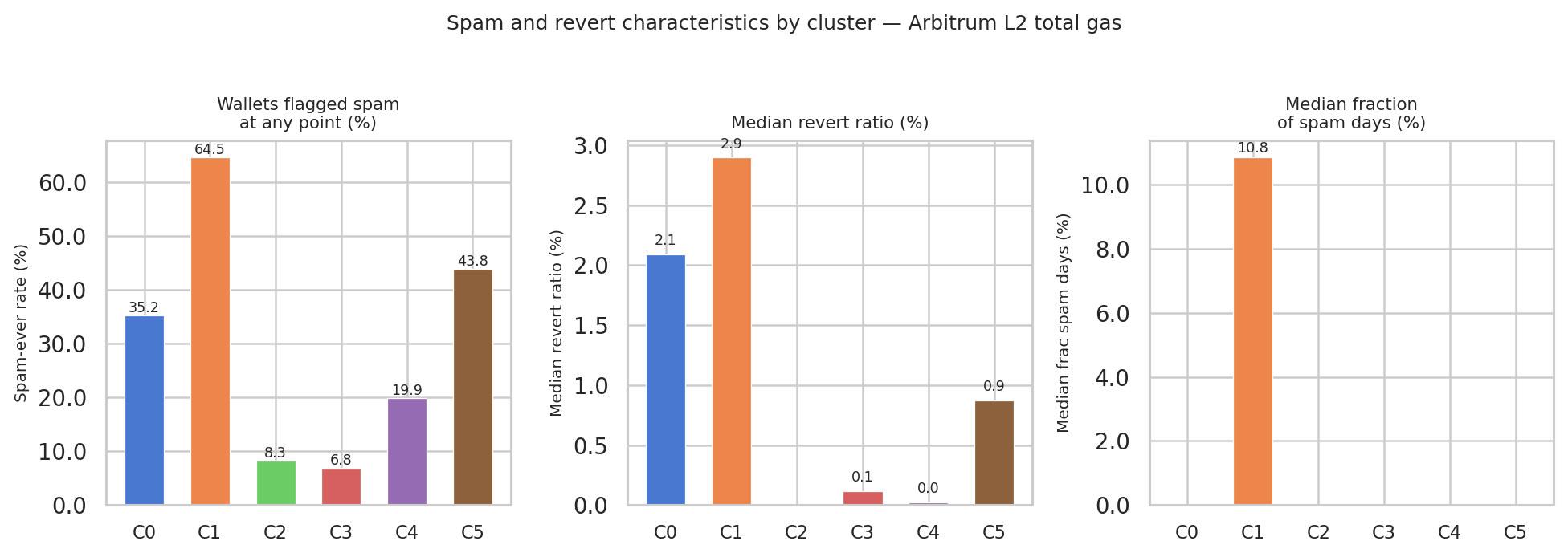}
  \caption{\Adv activity metrics by cluster for the \arb L2 total-gas
    analysis. Left: \adv-ever rate. Center: median revert ratio.
    Right: median fraction of \adv days.}
  \label{fig:adv-l2-total}
\end{figure}

\subsubsection*{L2 per-resource clusters}

C1 has the highest median revert ratio ($7.65\%$) and a non-zero \adv-day fraction
($3.1\%$); wallets in this cluster are both computation and read heavy
(Table~\ref{tab:clusters-l2-multidim}). C2 has a high \adv-ever rate ($49.5\%$)
with significantly higher current prevalence than C1, indicating ongoing rather than
historical \adv activity. C4 is the cleanest cluster ($3.5\%$ ever, zero revert,
zero \adv days), consistent with non-competitive retail wallets.

\begin{table}[t]
  \centering
  \caption{\Adv activity by cluster, \arb L2 per-resource analysis. Columns as in
    Table~\ref{tab:adv-l2-total}.}
  \label{tab:adv-l2-multidim}
  \small
  \begin{tabular}{lrrrr}
    \toprule
    Cluster & \makecell{\Adv-ever\\(\%)} & \makecell{Current\\(\%)}
      & \makecell{Med.\ revert\\(\%)} & \makecell{Med.\ \adv\\days (\%)} \\
    \midrule
    C0 & 25.3 &  2.7 & 0.05 & 0.0 \\
    C1 & 50.9 &  1.0 & 7.65 & 3.1 \\
    C2 & 49.5 & 24.4 & 1.74 & 0.0 \\
    C3 & 24.1 & 10.4 & 0.42 & 0.0 \\
    C4 &  3.5 &  0.8 & 0.00 & 0.0 \\
    C5 &  9.2 &  0.7 & 0.19 & 0.0 \\
    \bottomrule
  \end{tabular}
\end{table}

\section{Demeaned Binscatter Plots}
\label{app:binscatter}

The following figures show the two-way demeaned binscatter for L1 and L2,
confirming the negative demand slope in the partialled-out data.

\begin{figure}[t]
  \centering
  \includegraphics[width=0.72\linewidth]{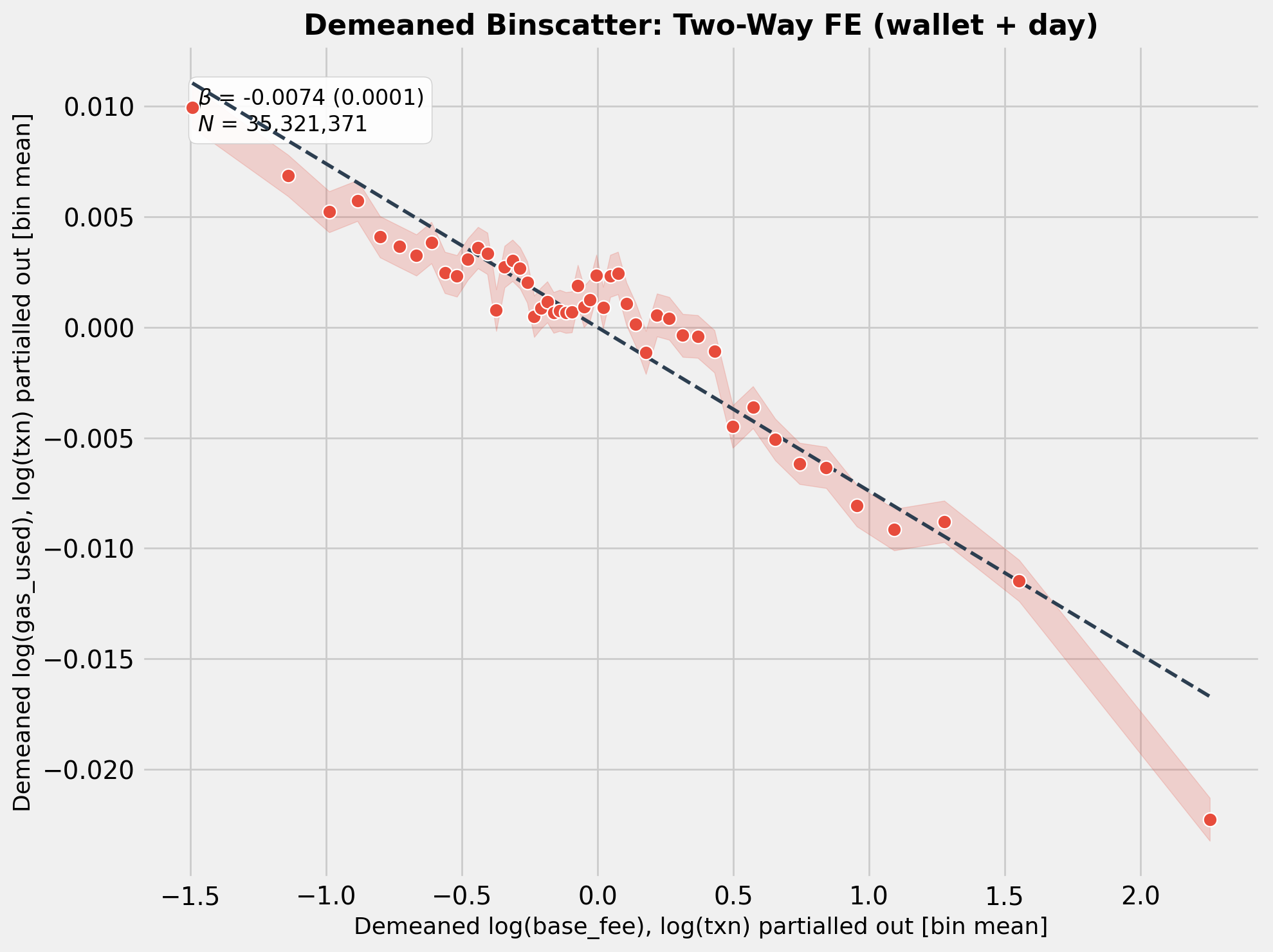}
  \caption{Two-way demeaned binscatter for L1 (Ethereum mainnet, daily FE).
    Negative slope after removing wallet and day fixed effects.}
  \label{fig:l1-binscatter}
\end{figure}

\begin{figure}[t]
  \centering
  \includegraphics[width=0.72\linewidth]{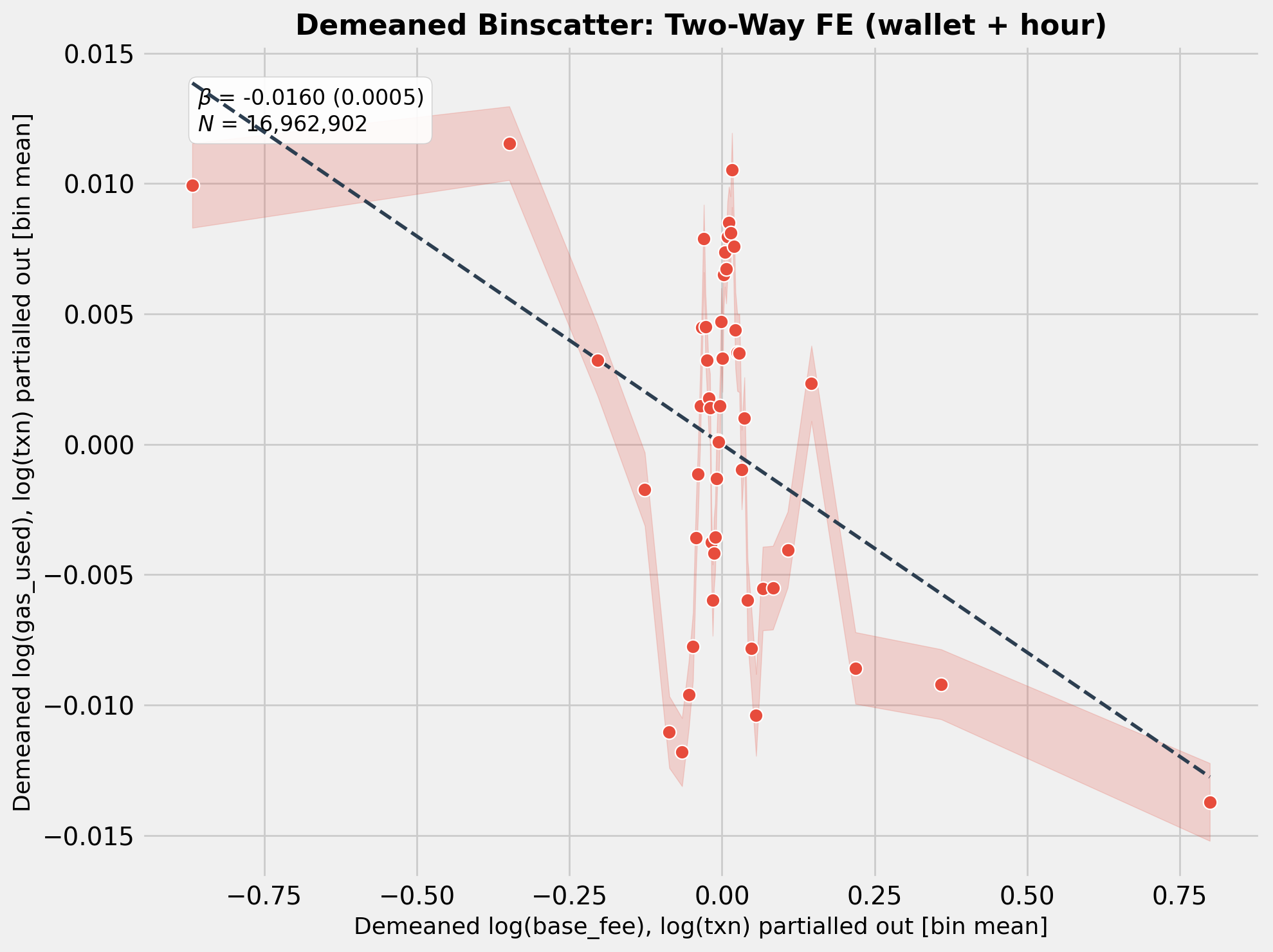}
  \caption{Two-way demeaned binscatter for L2 total-gas analysis (\arb,
    hourly FE, October 2025--April 2026).}
  \label{fig:l2-total-binscatter}
\end{figure}

\section{Per-Resource Selection by Fee Bin}
\label{app:selection}

Each panel shows the share of active wallet-hours with positive consumption of the
given resource dimension, by base-fee decile. Flat or rising shares indicate
fee-driven compositional selection toward heavier users; declining shares indicate
that some wallets cease using that resource at higher fees.

\begin{figure}[t]
  \centering
  \includegraphics[width=\linewidth]{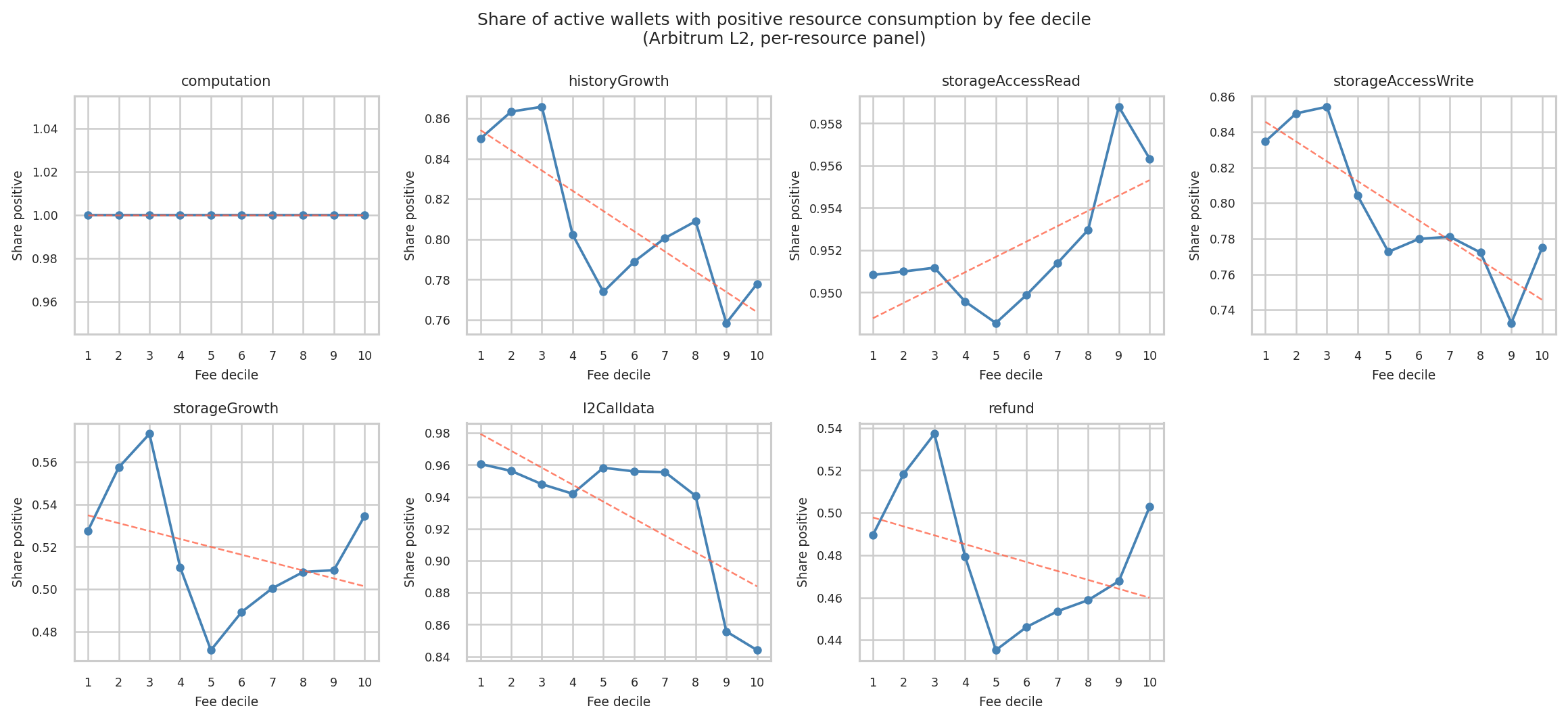}
  \caption{Share of active wallet-hours with positive resource consumption, by
    base-fee decile and dimension. Computation saturates at 1.0 across all deciles.
    Storage reads are flat (0.951--0.956). Storage writes, history growth, and
    calldata decline at higher fee levels. Storage growth and refunds are
    non-monotonic, consistent with their significantly negative IV estimates.}
  \label{fig:selection-by-fee-bin}
\end{figure}

\end{document}

%% file: aft_paper/paper/causal_fig.tex
\begin{figure}[t]
  \centering
  \begin{tikzpicture}[
      node distance = 2.6cm and 3.2cm,
      box/.style = {draw, rectangle, rounded corners = 3pt,
                    minimum width = 2.0cm, minimum height = 0.75cm,
                    align = center, font = \small},
      >=Latex
  ]
    \node[box] (z) {$z_{it}$\\\scriptsize lagged fee};
    \node[box, right = of z] (p) {$p_{it}$\\\scriptsize base fee};
    \node[box, right = of p] (g) {$g_{it}$\\\scriptsize gas usage};
    \node[box, above = 1.6cm of p, xshift = 1.6cm]
          (u) {$u_{it}$\\\scriptsize congestion};

    \draw[->, thick]
      (z) -- node[above, font=\footnotesize]{relevance} (p);

    \draw[->, thick]
      (p) -- node[above, font=\footnotesize]{$\beta$} (g);

    \draw[->, dashed, red!80!black, thick]
      (u) -- (p);
    \draw[->, dashed, red!80!black, thick]
      (u) -- (g);

    \draw[->, dashed, gray!70, thick]
      (z) to[out = -25, in = -155]
      node[below, font=\footnotesize, black]{$\times$ excluded} (g);

  \end{tikzpicture}
  \caption{Causal structure of the gas demand model.
    Congestion $u_{it}$ drives both the base fee $p_{it}$ (base fee update rule) and gas usage $g_{it}$ (through aggregate
    demand), inducing the endogeneity that biases OLS.
    The instrument $z_{it}$ (the wallet's lagged fee) satisfies relevance
    ($z_{it}$ predicts $p_{it}$) and the exclusion restriction ($z_{it}$ has
    no direct path to $g_{it}$, marked $\times$).
    Instrumental variables estimation uses only the variation in $p_{it}$
    attributable to $z_{it}$, which is uncorrelated with $u_{it}$.}
  \label{fig:dag}
\end{figure}

%% file: aft_paper/paper/feature_tab.tex
\begin{table}[t]
  \centering
  \caption{Clustering features. The top five are used in all analyses;
    the bottom four are added for the L2 per-resource clustering only.
    All features are standardised to zero mean and unit variance before
    clustering.}
  \label{tab:features}
  \small
  \begin{tabular}{lp{8.5cm}}
    \toprule
    Feature & What it captures \\
    \midrule
    \multicolumn{2}{l}{\textit{Behavioural features (all analyses)}} \\[2pt]
    $\log(\text{tx\_frequency})$   & Transaction rate per active hour; separates high-throughput operators from occasional users. \\[3pt]
    $\log(\text{gas/tx})$          & Mean gas per transaction; separates simple transfers from complex contract interactions. \\[3pt]
    $\log(\text{active\_hours})$   & Total hours with at least one transaction; separates persistent from episodic wallets. \\[3pt]
    $\text{gas\_cov}$              & Coefficient of variation ($\sigma/\mu$) of gas per transaction; low values indicate a fixed operation type, high values indicate volatile mixed activity. \\[3pt]
    $\text{high\_fee\_fraction}$   & Fraction of active hours above the median hourly fee; identifies wallets that transact preferentially during congestion. Also used as the exclusion restriction diagnostic (Section~4). \\[6pt]
    \multicolumn{2}{l}{\textit{Resource composition features (L2 per-resource only)}} \\[2pt]
    $\text{frac\_computation}$      & Share of gas from EVM opcode execution. \\[3pt]
    $\text{frac\_storageAccessRead}$& Share from reading existing storage slots. \\[3pt]
    $\text{frac\_l2Calldata}$       & Share from L2 calldata. \\[3pt]
    $\text{frac\_storageGrowth}$    & Share from creating new storage slots. \\
    \bottomrule
  \end{tabular}
\end{table}

%% file: aft_paper/paper/elbow_pictures.tex
\begin{figure}[t]
  \centering
  \begin{minipage}[t]{0.48\linewidth}
    \centering
    \includegraphics[width=\linewidth]{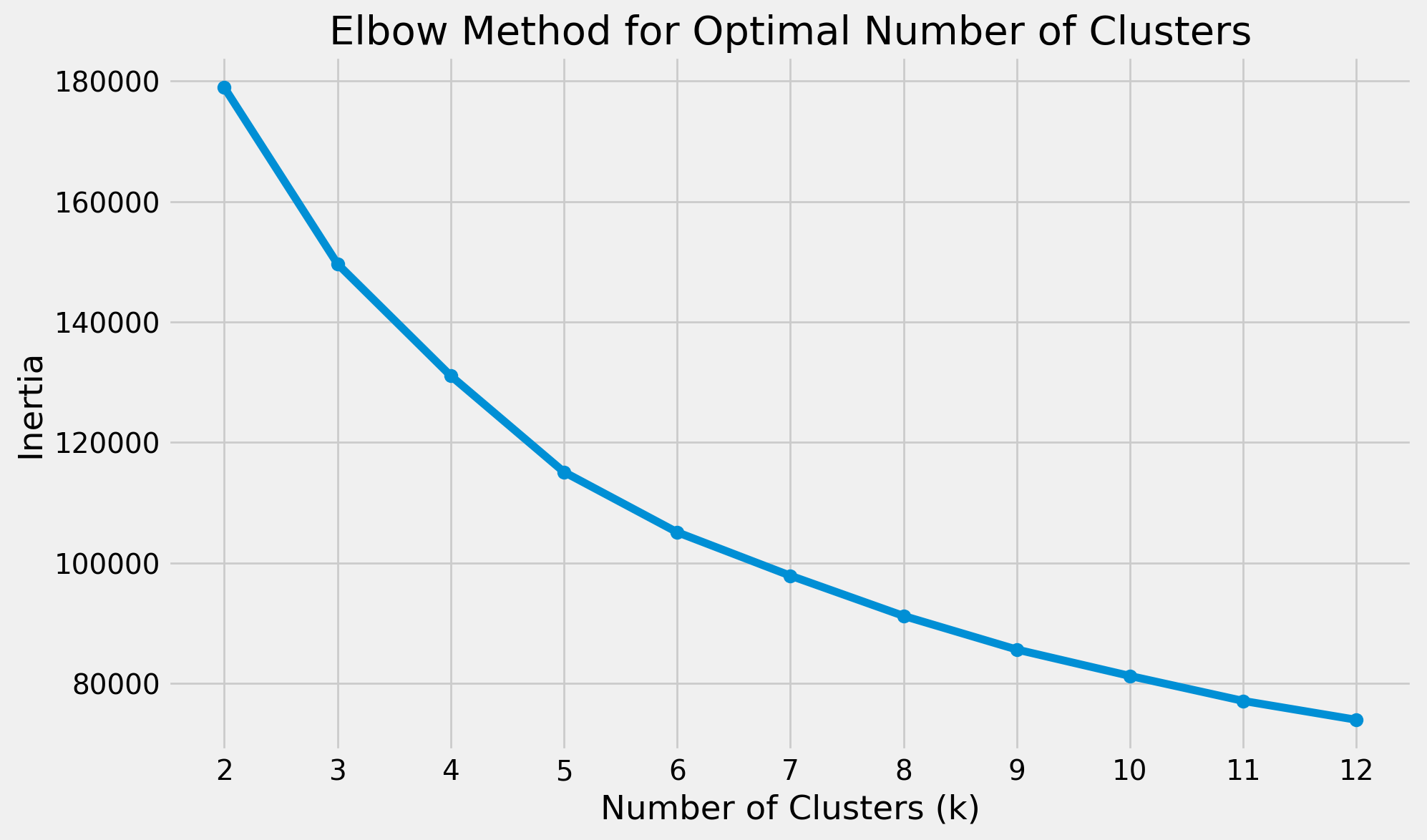}
    \captionof{figure}{Elbow plot for L1 (Ethereum) total-gas clustering.
      Within-cluster inertia as a function of $k$; the kink at $k = 6$
      identifies the chosen number of clusters.}
    \label{fig:elbow-l1}
  \end{minipage}
  \hfill
  \begin{minipage}[t]{0.48\linewidth}
    \centering
    \includegraphics[width=\linewidth]{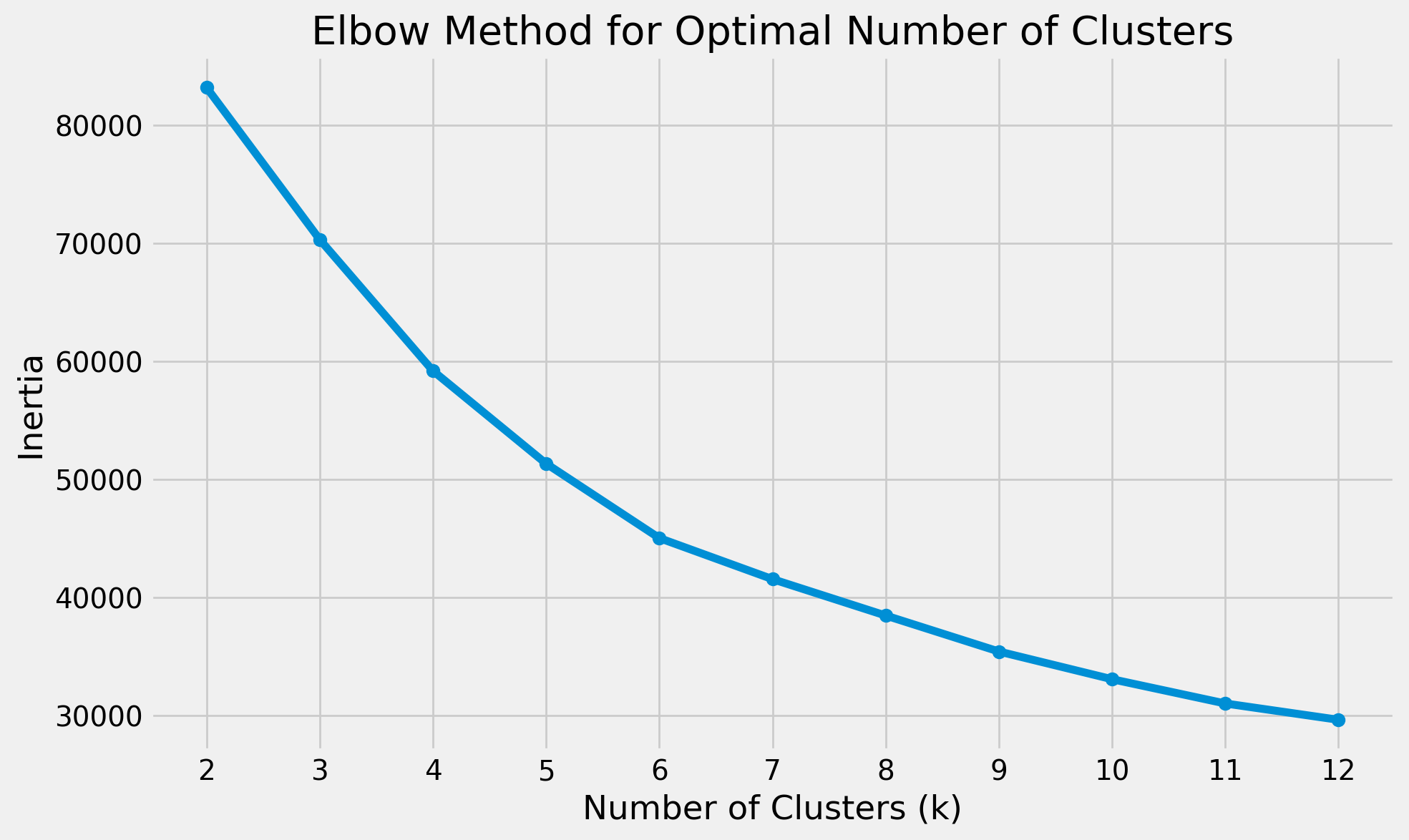}
    \captionof{figure}{Elbow plot for L2 (Arbitrum) total-gas clustering.
      Same criterion applied to the L2 panel; $k = 6$ selected.}
    \label{fig:elbow-l2}
  \end{minipage}
\end{figure}

\begin{figure}[t]
  \centering
  \includegraphics[width=0.5\linewidth]{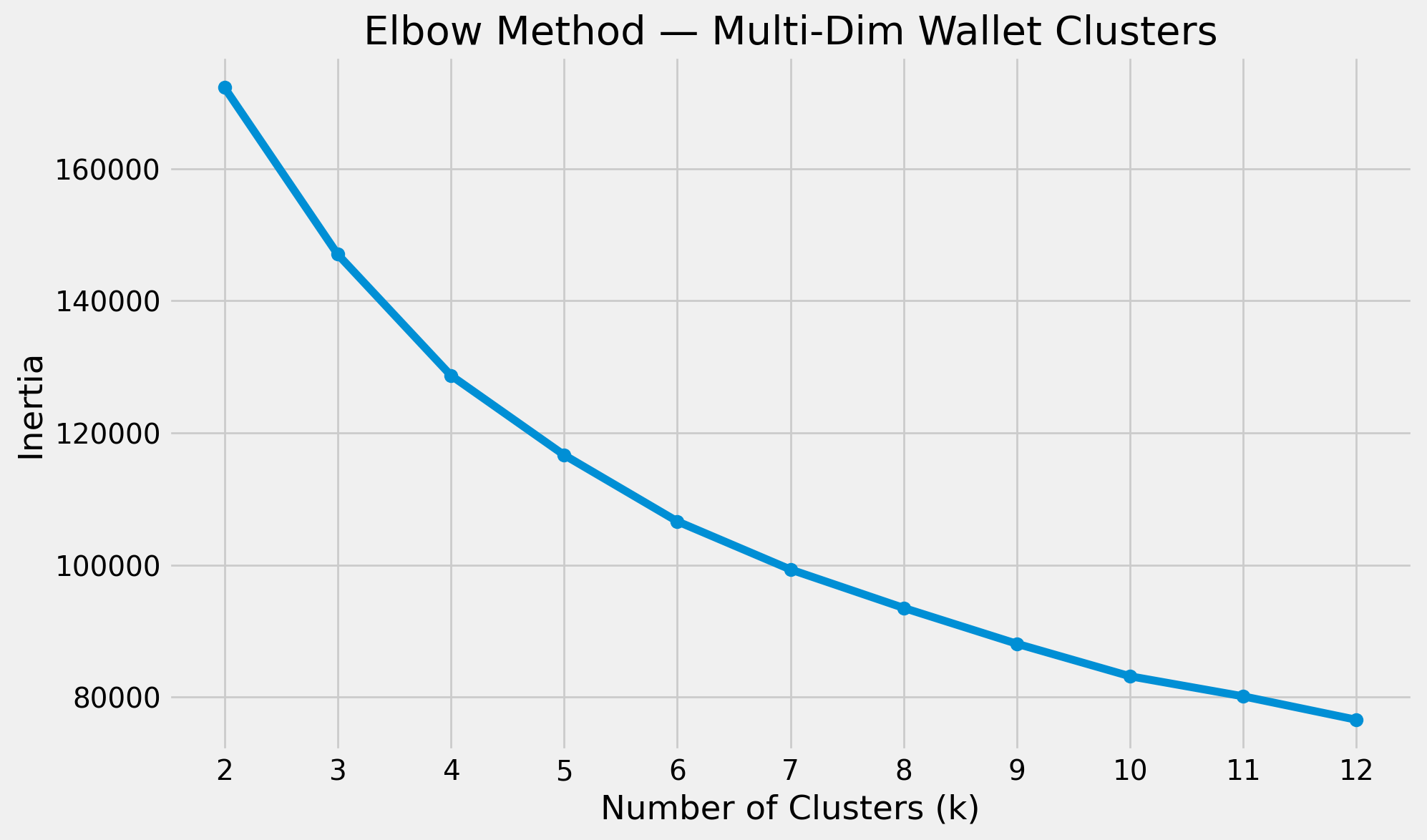}
  \caption{Elbow plot for L2 per-resource clustering on nine features.
    The kink is again at $k = 6$.}
  \label{fig:elbow-multidim}
\end{figure}

%% file: refs.bib
@article{roughgarden2021eip1559,
  author    = {Tim Roughgarden},
  title     = {{Transaction Fee Mechanism Design for the Ethereum Blockchain:
                An Economic Analysis of EIP-1559}},
  journal   = {Cryptoeconomic Systems},
  year      = {2021},
  url       = {https://arxiv.org/abs/2012.00854},
}

@inproceedings{roughgarden2021mechanism,
  author    = {Tim Roughgarden},
  title     = {{Transaction Fee Mechanism Design}},
  booktitle = {Proceedings of the 22nd ACM Conference on Economics and Computation (EC 2021)},
  year      = {2021},
  doi       = {10.1145/3465456.3467660},
}

@inproceedings{ferreira2021dynamic,
  author    = {Matheus V.X. Ferreira and Daniel J. Moroz and David C. Parkes and Mitchell Stern},
  title     = {Dynamic Posted-Price Mechanisms for the Blockchain Transaction-Fee Market},
  booktitle = {Proceedings of the 3rd ACM Conference on Advances in Financial Technologies},
  year      = {2021},
  doi       = {10.1145/3479722.3480993},
}

@inproceedings{chung2023,
  author    = {Hao Chung and Elaine Shi},
  title     = {Foundations of Transaction Fee Mechanism Design},
  booktitle = {Proceedings of the 2023 Annual ACM-SIAM Symposium on Discrete Algorithms (SODA 2023)},
  year      = {2023},
  url       = {https://arxiv.org/abs/2111.03151},
}

@article{leonardos2021,
  author    = {Stefanos Leonardos and Barnab{\'e} Monnot and Danijar Reijsbergen
               and Efstratios Skoulakis and Georgios Piliouras},
  title     = {{Dynamical Analysis of the EIP-1559 Ethereum Fee Market}},
  journal   = {CoRR},
  volume    = {abs/2102.10567},
  year      = {2021},
  url       = {https://arxiv.org/abs/2102.10567},
}

@inproceedings{reijsbergen2021,
  author    = {Danijar Reijsbergen and Stefanos Leonardos and Barnab{\'e} Monnot
               and Stratis Skoulakis and Georgios Piliouras},
  title     = {{Transaction Fees on a Honeymoon: Ethereum's EIP-1559 One Month Later}},
  booktitle = {2021 IEEE International Conference on Blockchain (Blockchain)},
  year      = {2021},
  doi       = {10.1109/Blockchain53845.2021.00049},
}

@inproceedings{liu2022,
  author    = {Yulin Liu and Yuxuan Lu and Kartik Nayak and Fan Zhang and Luyao Zhang and Yinhong Zhao},
  title     = {{Empirical Analysis of EIP-1559: Transaction Fees, Waiting Times, and Consensus Security}},
  booktitle = {Proceedings of the 2022 ACM SIGSAC Conference on Computer and Communications Security},
  year      = {2022},
  doi       = {10.1145/3548606.3559341},
}

@misc{silva2026elasticity,
  author       = {Maria Silva},
  title        = {{Empirical Analysis of Price Elasticities for Ethereum State and Burst Resources}},
  howpublished = {Ethereum Research},
  year         = {2026},
  url          = {https://ethresear.ch/t/empirical-analysis-of-price-elasticities-for-ethereum-state-and-burst-resources/24166},
}

@article{donmez2022,
  author    = {Anil Donmez and Alexander Karaivanov},
  title     = {{Transaction Fee Economics in the Ethereum Blockchain}},
  journal   = {Economic Inquiry},
  volume    = {60},
  number    = {1},
  pages     = {265--292},
  year      = {2022},
  doi       = {10.1111/ecin.13025},
}

@inproceedings{pierro2019,
  author    = {Giuseppe Antonio Pierro and Henrique Rocha},
  title     = {{The Influence Factors on Ethereum Transaction Fees}},
  booktitle = {2nd International Workshop on Emerging Trends in Software Engineering for Blockchain (WETSEB)},
  pages     = {24--31},
  year      = {2019},
  doi       = {10.1109/WETSEB.2019.00010},
}

@misc{monnot2022rollup,
  author       = {Barnab{\'e} Monnot},
  title        = {{Understanding Rollup Economics from First Principles}},
  howpublished = {Substack},
  year         = {2022},
  url          = {https://barnabe.substack.com/p/understanding-rollup-economics-from},
}

@misc{crapis2023rollup,
  author       = {Davide Crapis},
  title        = {{Rollups are Real: Rollup Economics 2.0}},
  howpublished = {Notion},
  year         = {2023},
  url          = {https://davidecrapis.notion.site/Rollups-are-Real-Rollup-Economics-2-0-2516079f62a745b598133a101ba5a3de},
}

@inproceedings{diamandis2023multidim,
  author    = {Theo Diamandis and Alex Evans and Tarun Chitra and Guillermo Angeris},
  title     = {{Dynamic Pricing for Non-Fungible Resources: Designing Multidimensional Blockchain Fee Markets}},
  booktitle = {5th Conference on Advances in Financial Technologies (AFT 2023)},
  series    = {LIPIcs},
  publisher = {Schloss Dagstuhl},
  year      = {2023},
  url       = {https://arxiv.org/abs/2208.07919},
}

@inproceedings{angeris2025optimal,
  author    = {Guillermo Angeris and Theo Diamandis and Ciamac C. Moallemi},
  title     = {{Multidimensional Blockchain Fees Are (Essentially) Optimal}},
  booktitle = {7th Conference on Advances in Financial Technologies (AFT 2025)},
  series    = {LIPIcs},
  publisher = {Schloss Dagstuhl},
  year      = {2025},
  url       = {https://arxiv.org/abs/2402.08661},
}

@inproceedings{victor2020clustering,
  author    = {Friedhelm Victor},
  title     = {{Address Clustering Heuristics for Ethereum}},
  booktitle = {Financial Cryptography and Data Security (FC) 2020},
  series    = {Lecture Notes in Computer Science},
  volume    = {12059},
  pages     = {617--633},
  publisher = {Springer},
  year      = {2020},
}

@inproceedings{beres2021profiling,
  author    = {Ferenc B{\'e}res and Istv{\'a}n Andr{\'a}s Seres and Andr{\'a}s A. Bencz{\'u}r and Mikerah Quintyne-Collins},
  title     = {{Blockchain Is Watching You: Profiling and Deanonymizing Ethereum Users}},
  booktitle = {IEEE International Conference on Decentralized Applications and Infrastructures (DAPPS)},
  pages     = {69--78},
  year      = {2021},
  url       = {https://arxiv.org/abs/2005.14051},
}

@article{hu2024bots,
  author    = {Hanzhi Hu and Tom Beer and Daniel Perez},
  title     = {{Detecting Financial Bots on the Ethereum Blockchain}},
  journal   = {CoRR},
  volume    = {abs/2403.19530},
  year      = {2024},
  url       = {https://arxiv.org/abs/2403.19530},
}

@article{borenstein2005realtime,
  author    = {Severin Borenstein},
  title     = {{The Long-Run Efficiency of Real-Time Electricity Pricing}},
  journal   = {The Energy Journal},
  volume    = {26},
  number    = {3},
  pages     = {93--116},
  year      = {2005},
}

@article{fabra2021elasticity,
  author    = {Natalia Fabra and David Rapson and Mar Reguant and Jingyuan Wang},
  title     = {{Estimating the Elasticity to Real-Time Pricing: Evidence from the Spanish Electricity Market}},
  journal   = {AEA Papers and Proceedings},
  volume    = {111},
  pages     = {425--429},
  year      = {2021},
  doi       = {10.1257/pandp.20211007},
}

@article{tiedemann2024iv,
  author    = {Silvana Tiedemann and Raffaele Sgarlato and Lion Hirth},
  title     = {{Price Elasticity of Electricity Demand: Using Instrumental Variable Regressions to Address Endogeneity and Autocorrelation of High-Frequency Time Series}},
  journal   = {Energy Economics},
  year      = {2024},
  url       = {https://arxiv.org/abs/2306.12863},
}

@article{frisch1933partial,
  title     = {Partial Time Regressions as Compared with Individual Trends},
  author    = {Frisch, Ragnar and Waugh, Frederick V.},
  journal   = {Econometrica},
  volume    = {1},
  number    = {4},
  pages     = {387--401},
  year      = {1933},
  doi       = {10.2307/1907330},
}

@article{lovell1963seasonal,
  title     = {Seasonal Adjustment of Economic Time Series and Multiple Regression Analysis},
  author    = {Lovell, Michael C.},
  journal   = {Journal of the American Statistical Association},
  volume    = {58},
  number    = {304},
  pages     = {993--1010},
  year      = {1963},
  doi       = {10.1080/01621459.1963.10480682},
}

@book{angrist2009,
  author    = {Joshua D. Angrist and J{\"o}rn-Steffen Pischke},
  title     = {{Mostly Harmless Econometrics: An Empiricist's Companion}},
  publisher = {Princeton University Press},
  year      = {2009},
}

@book{wooldridge2010,
  author    = {Jeffrey M. Wooldridge},
  title     = {{Econometric Analysis of Cross Section and Panel Data}},
  edition   = {2nd},
  publisher = {MIT Press},
  year      = {2010},
}

@article{angristkrueger1991,
  author    = {Joshua D. Angrist and Alan B. Krueger},
  title     = {{Does Compulsory School Attendance Affect Schooling and Earnings?}},
  journal   = {The Quarterly Journal of Economics},
  volume    = {106},
  number    = {4},
  pages     = {979--1014},
  year      = {1991},
  doi       = {10.2307/2937954},
}

@article{staiger1997,
  author    = {Douglas Staiger and James H. Stock},
  title     = {{Instrumental Variables Regression with Weak Instruments}},
  journal   = {Econometrica},
  volume    = {65},
  number    = {3},
  pages     = {557--586},
  year      = {1997},
  doi       = {10.2307/2171753},
}

@misc{buterin2019eip1559,
  author    = {Vitalik Buterin and Eric Conner and Rick Dudley and Matthew Slipper
               and Ian Norden and Abdelhamid Bakhta},
  title     = {{EIP-1559: Fee Market Change for ETH 1.0 Chain}},
  howpublished = {Ethereum Improvement Proposal},
  year      = {2019},
  url       = {https://eips.ethereum.org/EIPS/eip-1559},
}

@article{multi-dimensional,
  author       = {Nir Lavee and
                  Noam Nisan and
                  Mallesh M. Pai and
                  Max Resnick},
  title        = {Does Your Blockchain Need Multidimensional Transaction Fees?},
  journal      = {CoRR},
  volume       = {abs/2504.15438},
  year         = {2025},
  url          = {https://doi.org/10.48550/arXiv.2504.15438},
  doi          = {10.48550/ARXIV.2504.15438},
  eprinttype   = {arXiv},
  eprint       = {2504.15438},
  bibsource    = {dblp computer science bibliography, https://dblp.org}
}

@inproceedings{optimistic_mev,
  author       = {Ozan Solmaz and
                  Lioba Heimbach and
                  Yann Vonlanthen and
                  Roger Wattenhofer},
  editor       = {Zeta Avarikioti and
                  Nicolas Christin},
  title        = {Optimistic {MEV} in Ethereum Layer 2s: Why Blockspace Is Always in
                  Demand},
  booktitle    = {7th Conference on Advances in Financial Technologies, {AFT} 2025,
                  Pittsburgh, PA, USA, October 8-10, 2025},
  series       = {LIPIcs},
  pages        = {28:1--28:24},
  publisher    = {Schloss Dagstuhl - Leibniz-Zentrum f{\"{u}}r Informatik},
  year         = {2025},
  url          = {https://doi.org/10.4230/LIPIcs.AFT.2025.28},
  doi          = {10.4230/LIPICS.AFT.2025.28},
  bibsource    = {dblp computer science bibliography, https://dblp.org}
}

@article{blockspace_pressure,
  author       = {Wenhao Wang and
                  Aditya Saraf and
                  Lioba Heimbach and
                  Kushal Babel and
                  Fan Zhang},
  title        = {Blockspace Under Pressure: An Analysis of Spam {MEV} on High-Throughput
                  Blockchains},
  journal      = {CoRR},
  volume       = {abs/2604.00234},
  year         = {2026},
  url          = {https://doi.org/10.48550/arXiv.2604.00234},
  doi          = {10.48550/ARXIV.2604.00234},
  eprinttype   = {arXiv},
  eprint       = {2604.00234},
  bibsource    = {dblp computer science bibliography, https://dblp.org}
}

@article{timing_games,
  author       = {Bruno Mazorra and
                  Christoph Schlegel and
                  Akaki Mamageishvili},
  title        = {Timing Games: Probabilistic backrunning and spam},
  journal      = {CoRR},
  volume       = {abs/2602.22032},
  year         = {2026},
  url          = {https://doi.org/10.48550/arXiv.2602.22032},
  doi          = {10.48550/ARXIV.2602.22032},
  eprinttype   = {arXiv},
  eprint       = {2602.22032},
  bibsource    = {dblp computer science bibliography, https://dblp.org}
}

@article{one_vs_multi,
  author       = {Aggelos Kiayias and
                  Elias Koutsoupias and
                  Giorgos Panagiotakos and
                  Kyriaki Zioga},
  title        = {One-dimensional vs. Multi-dimensional Pricing in Blockchain Protocols},
  journal      = {CoRR},
  volume       = {abs/2506.13271},
  year         = {2025},
  url          = {https://doi.org/10.48550/arXiv.2506.13271},
  doi          = {10.48550/ARXIV.2506.13271},
  eprinttype   = {arXiv},
  eprint       = {2506.13271},
  bibsource    = {dblp computer science bibliography, https://dblp.org}
}

@inproceedings{optimal_dynamic_fees,
  author       = {Davide Crapis and
                  Ciamac C. Moallemi and
                  Shouqiao Wang},
  editor       = {Jeremy Clark and
                  Elaine Shi},
  title        = {Optimal Dynamic Fees for Blockchain Resources},
  booktitle    = {Financial Cryptography and Data Security - 28th International Conference,
                  {FC} 2024, Willemstad, Cura{\c{c}}ao, March 4-8, 2024, Revised Selected
                  Papers, Part {I}},
  series       = {Lecture Notes in Computer Science},
  pages        = {271--291},
  publisher    = {Springer},
  year         = {2024},
  url          = {https://doi.org/10.1007/978-3-031-78676-1\_15},
  doi          = {10.1007/978-3-031-78676-1\_15},
  bibsource    = {dblp computer science bibliography, https://dblp.org}
}

@inproceedings{l2_posting_on_l1,
  author       = {Shouqiao Wang and
                  Davide Crapis and
                  Ciamac C. Moallemi},
  editor       = {Christina Garman and
                  Pedro Moreno{-}Sanchez},
  title        = {A Framework for Combined Transaction Posting and Pricing for Layer
                  2 Blockchains},
  booktitle    = {Financial Cryptography and Data Security - 29th International Conference,
                  {FC} 2025, Miyakojima, Japan, April 14-18, 2025, Revised Selected
                  Papers, Part {I}},
  series       = {Lecture Notes in Computer Science},
  pages        = {56--72},
  publisher    = {Springer},
  year         = {2025},
  url          = {https://doi.org/10.1007/978-3-032-07024-1\_4},
  doi          = {10.1007/978-3-032-07024-1\_4},
  bibsource    = {dblp computer science bibliography, https://dblp.org}
}
